\documentclass[aps,prc,twocolumn,floatfix,showpacs,a4paper,nofootinbib,amsmath,amssymb]{revtex4-1}
\usepackage{graphicx}
\usepackage{color}
\usepackage{dcolumn}
\usepackage{bm}
\newcommand{\be}{\begin{equation}}
\newcommand{\ee}{\end{equation}}
\newcommand{\ba}{\begin{eqnarray}}
\newcommand{\ea}{\end{eqnarray}}
\newcommand{\bd}{\begin{displaymath}}
\newcommand{\ed}{\end{displaymath}}
\renewcommand{\vec}[1]{\mbox{\boldmath$#1$}}

\begin{document}

\title{$\Lambda$ Polarization in an Exact Rotating and Expanding fluid dynamical model for peripheral heavy ion reactions}
\author{Yilong Xie, Robert C. Glastad, L\'aszl\'o P. Csernai}
\affiliation{ Institute of Physics and Technology, University of Bergen,
Allegaten 55, 5007 Bergen, Norway }

\begin{abstract}
We calculate the  $\Lambda$ polarization in 
an exact analytical, rotating model based on parameters 
extracted from a high resolution (3+1)D Particle-in-Cell 
Relativistic hydrodynamics calculation. 
The polarization is attributed to effects from thermal vorticity 
and for the first time the effects of the radial and axial
acceleration are also studied separately.
\end{abstract}

\date{\today\ - v6}


\maketitle

%%%%%%%%%%%%%%%%%%%%%%%%%%%%%%%%%%%
\section{Introduction}
%%%%%%%%%%%%%%%%%%%%%%%%%%%%%%%%%%%

In high energy peripheral heavy ion collisions there 
is a substantial amount of initial angular momentum
directly after the Lorentz contracted nuclei 
penetrate each other. The formed Quark Gluon Plasma
locally equilibrates, the shear flow leads to local rotation,
i.e. vorticity, and then it expands, while its rotation slows down.

Due to the finite impact parameter, the initial stages (IS) 
have a non-vanishing angular momentum \cite{M1,M2}. 
For the initial
stages, effective models as the color glass condensate (CGC) 
or Glauber model are used. In general, we use
experimental data, to construct
a possible IS, at a given impact parameter for the  participant nucleons,
and their eccentricity.
Early studies neglected
effects arising from the non-vanishing angular 
momentum, but interest increased recently 
\cite{McI-Teo,McI2014,Nona15,Pang15}.

After many decades of refinements \cite{CGSTNK82,CFR84}, 
hydrodynamical modelling became the 
best to describe the middle stages of  
heavy ion collisions at relativistic energies. 
Thus, rotation and its consequences in peripheral collisions
were also studied in fluid dynamical
models \cite{hydro1,hydro2}.

We look at polarization in effects arising from thermal vorticity 
in the exact rotating and expanding model
\cite{CsN14}, where we are 
modeling an appropriate time-period of the collision
\cite{CsWCs14-2}. 
Special attention has been given to the collective motion, and to extract it
from observables which could confirm that such descriptions are
indeed plausible. 

We calibrate an exact rotating model based on
a (3+1)D fluid dynamical model, the relativistic particle-in-cell
method (PICR), to fine-tune the initial parameters of the rotating and
expanding fireball
\cite{CsWCs14-2}.

In Ref. \cite{CsV13} the differential Hanbury Brown and Twiss
(HBT)-method was used to detect rotation in heavy ion collisions. 

Without at least some viscosity and/or interaction one could not
generate rotation from the original shear flow. On the other hand
to develop instabilities or turbulence the viscosity should be
small, so that the ratio of shear viscosity to entropy density, 
$\eta/s$, should be of the order of 
$\hbar/4\pi k_B$, which can be achieved at the
phase transition between hadronic matter and QGP \cite{CKM}.

Thermal vorticity arises from the flow velocity field 
\cite{CMW13},
and the inverse temperature field present in
heavy ion collisions, and it arises mainly due to a non-vanishing angular
momentum in the initial stage.    

Fluctuations in the transverse plane 
can generate significant vorticity, 
but in peripheral collisions the initial shear flow 
leads to an order of magnitude larger vorticity.
\cite{CMW13}. This vorticity  may be further enhanced by 
the Kelvin Helmholtz Instability (KHI).

In our formalism, the dynamics of the system after local equilibration is
computed using the relativistic (3+1)D fluid dynamical model, PICR. 
This fluid dynamical (FD) computation with small viscosity 
shows enhanced collective rotation 
due to an evolving KHI. In Ref. \cite{WNC13} a simple analytic
model for this phenomenon is explored using a few material
properties: the surface tension between the colliding nuclei, the
viscosity, and the thickness of the flow layer. This enables a classical
potential flow approximation, in which one may study the dynamics
of an onsetting KHI.

A more recent calculation of the onset and effects of the KHI
is performed in Ref. \cite{CsWCs14-2}, in which the calibration of
the exact model takes place. Here, it is pointed out that this feature --
the enhancement of rotation -- is a dominant aspect of the (3+1)D fluid
dynamical model, but it is also seen in UrQMD \cite{GBL14}.

At high energy collisions,
we need an initial state model, which describes the dynamics
until local equilibration is reached. There are several options for
describing this pre-equilibrium dynamics, using Color Glass Condensate
(CGC) fields, parton (or hadron) kinetic theory, or one dimensional
Yang-Mills field (or flux tube) models
\cite{M1,M2}. In the (3+1)D PICR fluid dynamical model that 
we use as our guidance for the FD development, this last choice
is used.

 It is important to mention that for peripheral
collisions the initial shear and sometimes even the  angular 
momentum are neglected, while realistic initial state models
include these features
\cite{GBL14,Nona15,Pang15}.

>From the initial shear flow, in the  (3+1)D PICR fluid dynamical model 
the general rotation develops gradually in 1 to 2 fm/c time. 
Thus, the Exact model is applicable from this point of time on 
\cite{CsWCs14-2}. At the energies we discuss, by this time the matter
is in locally equilibrated QGP phase, and the local vorticity
 develops also. Due to the spin orbit interaction the
local vorticity and the spin of quarks equilibrate.
The essential part of the dynamical development of flow
(and other collective mechanical processes) takes place in the 
QGP phase, which is indicated by the Constituent Quark Number 
scaling of the flow harmonics. 

This most significant middle stage of the reaction can be
modeled by the "Exact" model
\cite{CsN14}.
The model is based on a set of scaling variables,
\be
(s_r,s_y)=\left(\frac{x^2+z^2}{R^2},\frac{y^2}{Y^2}\right),
\ee
in terms of the transverse and axial coordinates, $x$ $z$, and $y$;
and the characteristic radius $R$
and axial length, $Y$, parameters.
% Nucl. Phys. B {\bf 878}, 186 (2014)
The scaling parameter $s=s_r+s_y$ is also introduced, being 
the scaling variable as it appears in the 
thermodynamical relations. Here we have
interchanged the $y$- and $z$-axes to resonate 
with choice of axes in heavy ion collision 
literature, in which the reaction plane,
in which the system rotates, is spanned by 
$\vec{e}_x$ and $\vec{e}_z$, leaving the axis 
of rotation to be defined by $\vec{e}_y$.

Ref. \cite{CsWCs14-2} calibrates the parameters of the Exact model to the 
(3+1)D fluid dynamical model. The parameters are
extracted for experiments at 
$\sqrt{S_{NN}}=2.76 A\cdot TeV$ with impact 
parameter $b=0.7b_{Max}$
(See Fig. 1 and 2). In the (3+1)D PICR model, rotation 
may increase due to Kelvin-Helmholtz instability, whereas in
the Exact model -- and the later stages in the experiments
themselves -- rotation slows due to a transfer of energy to the
explosively increasing radial expansion of the system.
The Exact model, therefore, is suited to describe the period from
the equilibration of rotation up to
the freeze-out.

In \cite{CsN14} the solution for a 
flow of conserved number density, together with a
constant, temperature-independent compressibility, and 
a velocity field is described. 
Hence the solutions
take form, in cylindrical coordinates 
$(r,y,\phi)$, where $r=\sqrt{x^2+z^2}$ with an 
equation of motion, $\vec{\dot{r}}(t)= \vec{v}(\vec{r},t)$.
The Exact model assumes a linear velocity profile both in the
radial, $r$, and in the axial, $y$, directions. This leads to a
flow development where a fluid element starting from a point
$(r_0, y_0, \phi_0)$, and at a later time, $t$, reaches
the point
\ba
r(t)=r_{0}
\frac{R(t)}{R(t_0)},
\nonumber \\
y(t)=y_0\frac{Y(t)}{Y(t_0)},
\nonumber \\
\phi(t)=\phi_0+\int dt\, \omega(t),
\ea
showing explicitly how the solutions evolve in time, 
rotating and expanding fluid.  These equations
follow the time-evolution of the scaling variables 
in the radial and axial directions. This is a Cylindrically symmetric
setup with $X(t)=Z(t)$, $\sqrt{X^2(t)+Z^2(t)}=R(t)$ 
and, in general, $Y(t)\neq R(t)$.

We have chosen the $x, z$-plane as our plane of rotation, 
with $y$ being the axis of rotation. Our initial angular momentum, then,
points in the negative $y$-direction, with an absolute 
value of approximately $1.45\cdot 10^4 \hbar$. In an attempt to 
determine new observables, we propose a search 
for $\Lambda$ polarization. 
Although, the polarization could be described similarly for all
Fermions, we chose the $\Lambda$s, because it is straightforward 
to determine its polarization from its decay to $p$ and $\pi$, 
(where the $p$ is emitted into the direction of the polarization).
Actually such an experiment is already performed at RHIC, but
the results were averaged for $\Lambda$-emissions to all azimuths,
while we predict significant polarization for particles emitted
in the $\pm x$ direction in the reaction plane
\cite{BCsW13}.

 Our expectation is that this polarization, at
least in part, will be able to account for the polarization as
observed in peripheral regions in the first $10$-$15$ fm/c
following the impact in a heavy ion collision. 

In order to evaluate the polarization in the Exact model
we use the parametrization of the Exact model based on the
realistic (3+1)D  PICR fluid dynamical calculation
\cite{CsWCs14-2}, 
and use the vorticity calculated in the Exact model
with these parameters in Ref.
\cite{CsI15}.

%%%%%%%%%%%%%%%%%%%%%%%%%%%%%%%%%%%
\section{Freeze-out and Polarization}
%%%%%%%%%%%%%%%%%%%%%%%%%%%%%%%%%%%
\label{S-I}

Polarization of $\Lambda$s was subject to theoretical studies 
before, both in $p+p$ and in heavy ion reactions. In single
$p+p$ collisions forward production in small-transverse-momentum 
fragmentation was theoretically studied and also observed. These reactions
did result in much higher polarizations up to about 30\%
\cite{DM81}.

To apply this approach to heavy ion collisions is a complex
theoretical problem because several microscopic processes
can contribute to polarization and these can be combined with
different hadron formation mechanisms
\cite{LW05,ACHM02}. 
In Ref. \cite{LW05} it was contemplated that 
the final heavy ion results are dependent on the hadronization
mechanisms,  and the effect of the decay products of the 
polarized hyperons on the $v_2$ flow harmonics,  $v_2$, were studied.
Ref.
\cite{ACHM02}
has also studied the sensitivity of $\Lambda$ production on the
coalescence or recombination mechanisms of the hadron formation.

As the previous works discussed a wide variety and complexity
of the microscopic description of hadronization and the resulting
polarization, we have followed a simpler statistical picture,
based on some simple assumptions of a dilute gas of particles,
on the "Relativistic distribution function for particles
with spin at local thermodynamical equilibrium"
\cite{BeChZaGr13}.

This work does not address the mechanisms of hadronization
and the change of polarization during this process. It also barely discusses 
the equilibrium between particle polarization and local rotation
in thermal equilibrium for dilute gases.  Thus, this approach
is primarily applicable to the final hadronic matter.

We follow the same reaction mechanism as used in all
(3+1)D PICR publications since 2001. We do not assume a 3 stage
fluid dynamical process in QGP phase, mixed phase and hadronic
phase because the fastest adiabatic development in the
mixed phase would take 30-50 fm/c
\cite{CsK92}. 
Such a long expansion time would contradict
all two particle correlation measurements showing a
size and timespan at FO of less than 10 fm. Furthermore
it would also contradict to the observed Constituent Quark Number
Scaling and to the observed large $\bar{\Omega}$ abundance.
The only way out of these problems is supercooling in the 
QGP phase, followed by rapid hadronization
\cite{CC94,CM95},
and almost immediate freeze-out.

Thus in the PICR fluid dynamical calculations we discuss 
exclusively the QGP phase, even for supercooled QGP.
Based on the mechanical equilibrium, evidenced by the
Constituent quark number scaling, we have reason to assume that
during the FD evolution there is ample time to
equipartition the local rotation among all degrees of freedom
in QGP, due to the spin-orbit interaction. As this is
a strongly interacting form of matter the kinetic approximation 
as a dilute gas is not necessarily applicable, and the energy
momentum and local angular momentum should also be carried by the 
fields.  
\footnote{If we would consider only three valence
quarks in kinetic equilibrium according to 
\cite{BeChZaGr13},
then the polarization of a coalesced 
Baryon would be $\Pi_B \sim (\Pi_q)^3$, which would not be measurable.}
We have to assume that the rapid hadronization 
maintains equipartition among all degrees of freedom
carrying angular momentum. So, based on this assumption
we use the approach of 
\cite{BeChZaGr13}.

Actually the same applies the statistical and thermal 
equilibrium among (most of) the abundances of final hadron species.
This can be understood based on the fact that the
statistical factors are the same in rapid formation of 
hadrons as in thermal equilibrium.

We use the same assumptions for the Exact fluid dynamical model
as we used for the (3+1)D PICR fluid dynamics.
Based on the above, in the Exact model  the
energy weighted thermal vorticity was calculated
\cite{CsI15}. We explored
the total energy of the system and the energy of expansion, 
rotation, and internal energy components and their time dependence.
We observed the
transfer of energy from rotation to expansion, hence 
the rotation slows as the system expands until the freeze-out.

According to the quantum-field-theoretical approach 
\cite{BeChZaGr13},
the expectation value of $\Lambda$ polarization in 
an inverse temperature field,
$\beta^\mu (x)=u^\mu(x)/T(x)$, is
\be
\langle \Pi_\mu (x,p) \rangle =
\frac{1}{8}\epsilon_{\mu\rho\sigma\tau}(1-n_F)
\partial^\rho\beta^\sigma(x)\frac{p^\tau}{m},
\ee
where $\epsilon_{\mu \rho \sigma \tau}$ is the completely 
antisymmetric Levi-Civita symbol, $n_F$ is the Fermi-J\"uttner distribution
for spin-1/2 particles ($(1-n_F)$ is the Pauli blocking factor), and $p$
is the $\Lambda$ four-momentum.  We integrate this over some volume, and
ultimately over all of space, weighted by the number density, normalized by
number of particles in that volume, leaving a momentum-dependent polarization
four-vector in the participant frame of reference

\be
\Pi_\mu(p) = \hbar \epsilon_{\mu \sigma \rho \tau}
\frac{p^\tau}{8m} \frac
{\int d\Sigma_\lambda p^\lambda n_F(x,p) (1-n_F(x,p)) 
\partial^\rho \beta^\sigma}
{\int d\Sigma_\lambda p^\lambda n_F(x,p)} \ .
\ee

Note that, as opposed to electromagnetic phenomena, 
in which particle and anti-particle will have anti-aligned polarization
vectors, here it is shown that $\Lambda$ and 
$\bar{\Lambda}$ polarizations are aligned in 
vorticious thermal  flow fields.

While the average values of polarization may be as 
low as $1$-$2\%$, consistent with RHIC bounds, 
in some regions of momentum
space we see a larger polarization, about $5\%$ for 
momenta in the transverse plane and up to a momentum of $3\ GeV/c$. 
Kelvin-Helmholtz instabilities
may further enhance rotation, hence the thermal vorticity, defined as
\be
\overline{\omega}_{\mu \nu} (x)=
\frac{1}{2}(\partial_\nu\beta_\mu-\partial_\mu\beta_\nu), 
\ee
and thereby the signal strength increases by $10$-$20\%$. At LHC 
energies, there may be $5\%\ \Lambda$ polarization due
to the corona effect, single nucleon-nucleon collisions 
occurring outside of the reaction zone of the collision itself.
So attempts
should be made to further the understanding of this 
background, and remove it from measurements in order to further
isolate the $\Lambda$ polarization as it arises 
from the collision itself.

The $\Lambda$ polarization is determined by measuring the
angular distribution of the decay protons in the $\Lambda$'s rest 
frame. In this frame 
the $\Lambda$ polarization is $\vec \Pi_0(\vec p)$, which can be 
obtained by Lorentz boosting the polarization $\vec \Pi(\vec p)$
from the participant frame to the  $\Lambda$'s rest frame, 
\cite{BCsW13},
\be
\vec{\Pi}_0(\vec{p})=
\vec{\Pi}( p )-\frac
{\vec{p}}
{p^0 (p^0 + m)}
\vec{\Pi}( p ) \cdot \vec{p} \ ,
\label{Pi0}
\ee
where $(p^0, \vec p)$ is the $\Lambda$'s four-momentum and 
$m$ its mass.

Based on this equation we see that in order to maximize 
polarization, we need to choose momenta for the $\Lambda$ such that
they lie in the reaction plane, 
hence we fix $\vec{p}$ in the positive $x$-direction.

\begin{figure}[h]  %%%%%%%%%%
\begin{center}
\resizebox{0.9\columnwidth}{!}
{\includegraphics{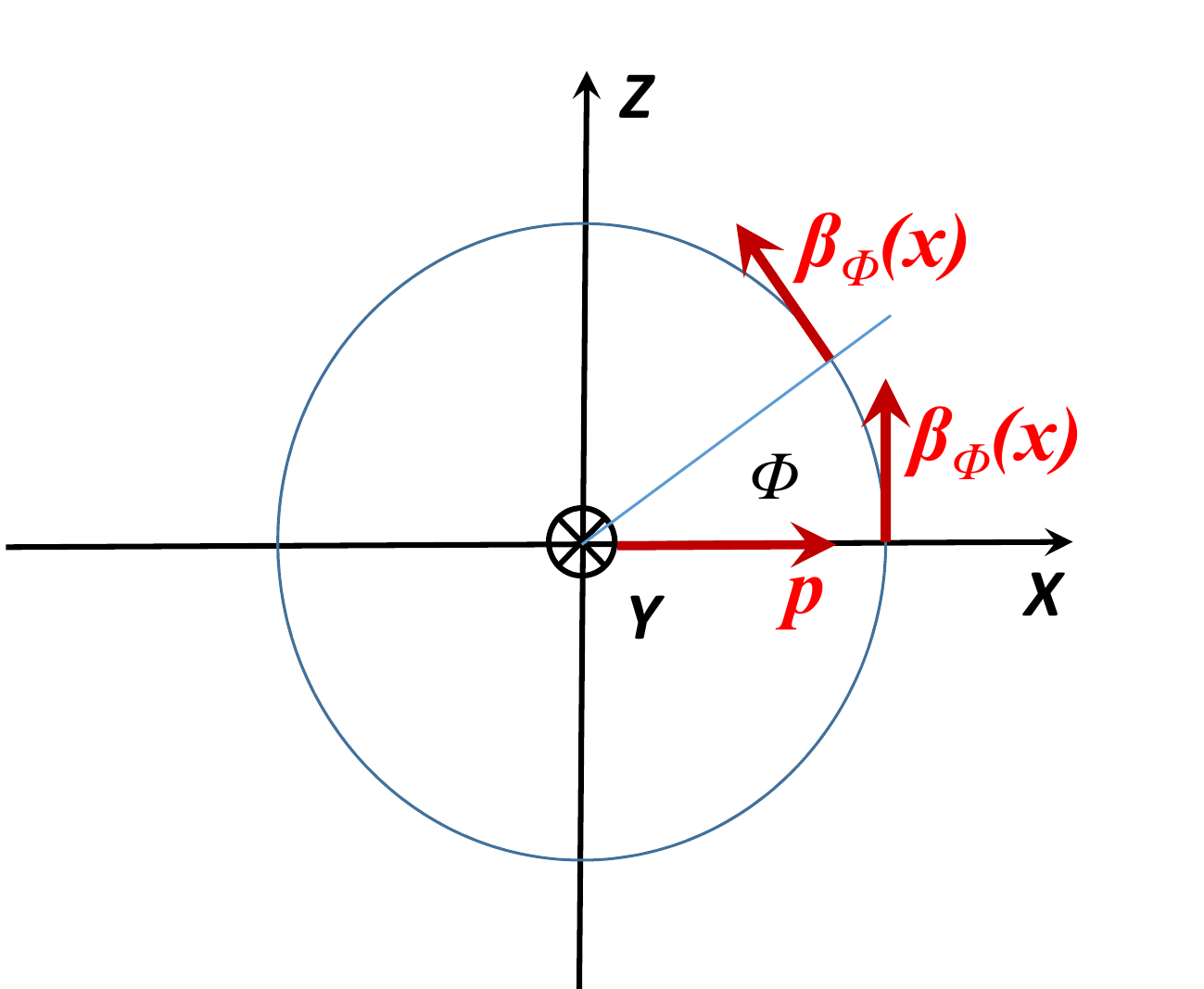}}
\caption{
(color online) 
The direction of axes, as well as the momentum, $\vec p$, and 
flow, $\vec \beta$, vectors. The azimuth angle is measured from 
the direction of the $\vec p$-vector, i.e. from the $x$-axis.
}
\label{Fig-xz}
\end{center}
\end{figure}
%%%%%%%%%%%%%%%%%%%%%%%%%%%%%%%%%%%

%%%%%%%%%%%%%%%%%%%%%%%%%%%%%%%%%%%
\section{Solution for the $\Lambda$ Polarization}
%%%%%%%%%%%%%%%%%%%%%%%%%%%%%%%%%%%
\label{S-II}

As the $\Lambda$ is transversely polarized, $\Pi^\mu p_\mu = 0$, one
can confine himself to the spatial part of $\Pi^\mu$.
The simplified spatial part of polarization vector is:

\ba
\vec{\Pi}(p) &=&
\frac{\hbar\epsilon}{8m}
\frac{\int \,d V n_F(x,p)\, (\nabla \times\vec{\beta})}
{\int \,d V n_F(x,p) }
\nonumber\\
&& + \frac{\hbar \vec p}{8m}
\times
\frac{\int \,d V n_F(x,p)\, (\partial_t \vec{\beta} + \nabla \beta^0)}
{\int \,d V n_F(x,p) }\ .
\label{Pol-1}
\ea
where $n_F(x,p)$ is the phase space distribution ofthe $\Lambda$s.
In a previous calculation \cite{BCsW13}, the $p$ dependence of  $n_F$,
was considered negligible in the integral and the time derivative and
gradient terms were also assumed to be smaller. The present calculation
shows that in general these terms are not negligible and that which terms are
dominant depends on the particular conditions.

We adopt the parametrization of the model from Ref. \cite{CsI15},
with the initial conditions
$R_0$ = 2.5 fm,  $Y_0$ = 4.0 fm,
$\dot R_0$ = 0.20 c,  $\dot Y_0$ = 0.25 c,
$\omega_0$ = 0.1 c/fm,
$\kappa = 3/2$,  $T_0$ = 300 MeV.
For this configuration $E_{tot}=576$ MeV/nucl.

%%%%%%%%%%%%%%%%%%%%%%%%%%%%%%%%%%%
\subsection{The denominator}
%%%%%%%%%%%%%%%%%%%%%%%%%%%%%%%%%%%

We first perform the integral in the denominator:
\be
A(p)\equiv \int \,d V \, n_F
 = \int\limits_0^R r \,d r \int\limits_{-Y}^{+Y} \,d y
 \int\limits_0^{2\pi} \,d \phi\ n_F(x,p) \ .
 \label{e2}
\ee

According to Eq. (3) in Ref. \cite{CsI15} in terms of the scaling 
variable, $s$, we have:
\ba
n &=& n_0\frac{V_0}{V}\nu(s) \ ,  \\
\nu(s) &=&
\frac{1}{\tau(s)} \exp\left(-\frac12 \int_0^s \frac{du}{\tau(u)}\right)
\nonumber\\
 &=& 1 \cdot \exp\left(-\frac12 \int_0^s du\right) \ ,
\ea
where the simplifying choice of $\tau(s)=1$ is used in the last step.
Therefore:
\be
n(s) = n_0\frac{V_0}{V}e^{-\frac12 s}\ .
\ee

 The EoS is assumed to be:
$\epsilon (s)=\kappa T(t)n(s)$
and the energy density $\epsilon(s)$ is calculated as
in Eq. (29)  in Ref. \cite{CsI15},
therefore:
\ba
n(s) &=& \frac{\epsilon}{\kappa T(t)} 
%\nonumber\\
%     &=& n_0 \frac{V_0}{V} \frac{T_0}{T}
%             \left(\frac{V_0}{V}\right)^{\frac{1}{\kappa}}\, e^{-\frac s2} 
%\nonumber\\
%     &=& n_0 \frac{T_0}{T} \left(\frac{V_0}{V}\right)^{1+1/\kappa}\,
%	 e^{-\frac{s_y}{2}} e^{-\frac{s_\rho}{2}} 
%\nonumber\\  &=&
= \frac{C_N}{\kappa T} e^{-\frac{s_y}{2}} e^{-\frac{s_\rho}{2}} \ ,
\label{e9}
\ea
where $C_N=\kappa n_0 T_0(\frac{V_0}{V})^{1+1/\kappa}$.

>From Ref. \cite{BCsW13}, the Fermi-J\"uttner distribution is:
\be
n_F(x,p) = \frac{1}{e^{p^\mu \beta_\mu-\xi}+1}
   \approx \frac{1}{e^{p^\mu \beta_\mu-\xi}}
         = \frac{e^{\mu/T}}{e^{p^\mu \beta_\mu}} \ ,
\ee
where the $\xi=\mu /T$, and $\mu$ is the chemical potential.
The thermal flow velocity, 
$\beta^\mu(x) \equiv u^\mu(x)/T$, is different 
at different space-time points $x$.

The invariant scalar density for the J\"unttner distribution is:
\be
n=\frac{4\pi m^2 K_2(m/T)}{(2\pi\hbar)^3}e^{\mu /T}
 =\frac{e^{\mu /T}}{C_0}
\ee
where the $C_0^{-1}={4\pi m^2 T K_2(m/T)}/{(2\pi\hbar)^3}$.
With $C_0$ and $n(s)=n$, the  Fermi-J\"uttner distribution can be written as:
\be
n_F(x,p)
= \frac{e^{\mu/T}}{e^{p^\mu \beta_\mu}}
= \frac{C_0 n(s)}{e^{p^\mu \beta_\mu}} \ .
\label{e12}
\ee
Now we introduce cylindrical coordinates for the location in the
configuration place $x = (r, y, \phi)$, and using the scaling
expansion model \cite{CsN14,CsWCs14-2} with the scaling variables
$s\,, s_r\,, s_y$.
Now, substituting Eqs. (\ref{e9},\ref{e12})
into the denominator of $\Pi(p)$, and parametrizing the range
of integrations as in \cite{CsI15}
one obtains:
\ba
A(p)&=&
%\int  dV\, n_F(p,s) 
%\nonumber\\
%= \frac{C_N C_0}{\kappa T} && \int r\, dr \int dy
%      \int d\phi\  n(s)\ e^{-p^\mu\beta_\mu}
%\nonumber\\ =
 \frac{C_N C_0}{\kappa T}\!\!
   \int^{aY}_{-aY}\!\!\!\!\!\!   dy\, \exp\left(-\frac{y^2}{2Y^2}\right)
   \int^{bR}_{0}\!\!\!\!\!\!\! r dr\, \exp\left(-\frac{r^2}{2R^2}\right)
\nonumber\\
   & & \qquad \times \int^{2\pi}_0 d\phi\, e^{-p^\mu \beta_\mu} \ .
\label{e15}
\ea

% In the above equation, we first integrate with respect to $\phi$.

%Generally,
%the position of the integration point in cylindrical coordinates is:
%$
%\vec{x} = (r, y, \phi) =
%$r \vec{e}_r + y \vec{e}_y + \vec{e}_{\phi}
%$.
%The spatial part of momentum vector, $p^\mu=(p^0,\vec{p})$,
%in cylindrical coordinates is:
%$
%\vec{p(x)} = (p_r, p_y, p_\phi) =
%p_r \vec{e}_r + p_y \vec{e}_y + p_\phi \vec{e}_{\phi}
%$, 
%and similarly:
%$
%\vec{\beta}(x) = (\beta_r, \beta_y, \beta_\phi )
%=\beta_r \vec{e}_r + \beta_y \vec{e}_y+\beta_\phi \vec{e}_{\phi}
%$.
The scalar product in cylindrical coordinates takes the form
$p^\mu\beta_\mu=(p^0,\vec{p})(\beta_0,\vec{\beta})
= p^0\beta_0-\vec{p\,\beta} 
= p^0\beta_0 - p_r \beta_r - p_y \beta_y - p_\phi \beta_\phi$.

In our integral the $p^\mu$ is given or 'fixed'
as the argument of $\vec{\Pi}(p)$,
while the $\vec \beta=\vec \beta(x)$ is changing. The integration with
respect to $\phi$ starts from the direction of the $\vec p$-vector.
According to the Eq. (5) in \cite{CsI15}:\\
$
\vec{v}=v_r \vec{e}_r+v_\phi \vec{e}_\phi+v_y \vec{e}_y =
\frac{\dot R}{R}r \vec{e}_r + \omega r \vec{e}_\phi +
\frac{\dot Y}{Y}y \vec{e}_y 
$, and
$
\vec{\beta}=u^i/T
=\gamma \vec{v}/T
$.
Thus in the integral for  $\phi$ we exploit the fact that in the Exact model
the radial, $r$, and axial, $y$, components of the thermal velocity,
$\vec \beta$, do not depend on $\phi$, 
while the tangential component
does not depend on $y$, i.e. $\beta_\phi=\gamma r\, \omega/T$, 
but its direction is changing with
respect to the direction of $\vec p$. As the integral is over the whole
$2\pi$ angle we can start it at any point of $\phi$, so we start it 
from the externally given $\vec p$-direction.
Consequently, with this choice of the $x$-axis, 
$\vec{p} = (p_r, p_y, 0)$, and $p_z = p_\phi = 0$. 
In this azimuthally symmetric, exact model it is sufficient
to calculate $\Pi(\vec p)$ for one direction of $\vec p$ in the 
$[x,z]$-plane.

%For the volume integration we still have to integrate over the 
%azimuthal direction $\phi$. The $\vec p \cdot \vec \beta$ product
%will be the same for every azimuthal angle as at the $x$-axis.

The direction of the thermal flow velocity, $\vec \beta$,
is tangential to the direction $\phi$, i.e. it points to the
$\vec e_{\phi+\pi/2}$-direction.
Thus the scalar product is: 
$$
\vec{p} \cdot \vec \beta(r,\, y,\phi) =
|p_x| \beta_r \cos{(\phi)}+p_y \beta_y + 
|p_x| \beta_\phi \cos{\left(\phi{+}\frac{\pi}{2}\right)} \ ,
$$
where $\phi$ is the azimuth angle of the position   
around the $y$, rotation axis,
counted starting from the $x$-axis.
%We have chosen the coordinate axes in a way that
%the $\vec{p}$ is orthogonal to the $z$ axis, $p_z=0$.  So,  
%the integral starts from $\phi = 0$, when $\vec{p}$ and 
%$\vec \beta_\phi$ are orthogonal. 
See Fig. \ref{Fig-xz}.

So, inserting the last expression for $p^\mu \beta_\mu$ into the last 
term of the integral Eq. (\ref{e15}),
the integral with respect to $\phi$ will take the form:
\ba
\int^{2\pi}_0 \!\!\!\!\!\! d\phi\, && e^{-p^\mu \beta_\mu} = 
%
% e^{-\gamma p^0/T} \, e^{p_y \beta_y} \times
%  \nonumber \\
%  && \int^{2\pi}_0 \!\!\!\!\!\! d\phi\, 
% e^{ |p_x| \beta_r \cos(\phi) - 
% |p_x| \beta_\phi \sin(\phi)} 
%\nonumber \\ =
  \int^{\pi}_{-\pi} \!\!\!\!\! d\phi\, e^{a \cos(\phi) - b \sin(\phi)}
%\nonumber \\
 = 2 \pi I_0\! \left( \sqrt{a^2{+}b^2} \right),
\nonumber \\
\label{eab}
\ea
where 
$a = |p_x| \beta_r   =|p_x| \gamma \dot R r/TR$  and
$b = |p_x| \beta_\phi=|p_x| \gamma r\, \omega/T$, and we used integral no.
3.338(4) in \cite{AlZ07}. 
If we define
$$c_3
=\sqrt{\left(\frac{p_x \gamma \dot R} {TR}\right)^2
     + \left(\frac{p_x \gamma \, \omega}{T}\right)^2}
= \frac{|p_x| \gamma}{T} \sqrt{(\dot R /R)^2+\omega^2}
\; ,
$$ 
then $\sqrt{a^2+b^2}=c_3 r$, and:
\ba
\int^{2\pi}_0 \!\!\!\!\!\! d\phi\, e^{-p^\mu \beta_\mu} 
=  e^{-\gamma p^0/T} \, e^{p_y \beta_y} \times 
2 \pi I_0 \left(c_3 r \right).
\label{e18}
\ea

Now, substituting this back into Eq. (\ref{e15}):
\ba
A(p) &=& \int dV\ n_F(p,s) =
\nonumber\\
%=\frac{C_N C_0}{\kappa T} &&
%\int\limits^{aY}_{-aY}dy e^{-\frac{y^2}{2Y^2}}
%\int\limits^{bR}_{0} r\, dr e^{-\frac{r^2}{2R^2}}
%\int\limits^{2\pi}_0 d\phi e^{-p^\mu \beta_\mu}
%\nonumber \\ =
\frac{C_N C_0}{\kappa T} &&
\int\limits^{aY}_{-aY}dr_y
\int\limits^{bR}_0 r\, dr \; 
\exp\Big(-\frac{y^2}{2Y^2}-\frac{r^2}{2R^2}\Big)
\nonumber \\
\times&&e^{-\gamma p^0/T} \, e^{p_y \beta_y} \cdot 
2 \pi I_0 \left(c_3\, r \right)\ .
\ea
Now we may use the same simplifying non-relativistic assumption as
in Eq. (5) of Ref. \cite{CsI15}, i.e. 
we approximate $u^\mu$ by $v^\mu$ as
$
\vec{v}=v_r \vec{e}_r  + v_y \vec{e}_y + v_\phi \vec{e}_\phi =
\frac{\dot R}{R}r \vec{e}_r  + \frac{\dot Y}{Y} y \vec{e}_y 
     + \omega r \vec{e}_\phi
$,
and thus $\gamma=1$. It follows then:
\ba
A(p) &=& \int dV\ n_F(p,s)
\nonumber \\
%= \frac{C_N C_0}{\kappa T} &&
%2\pi e^{-p^0/T}
%\int\limits^{aY}_{-aY}dy
%\exp\Big(\frac{p_y y \dot Y}{TY}-
%\frac{y^2}{2Y^2}\Big)
%\nonumber \\
%\times&& 
%\int\limits^{bR}_{0}\ r\ I_0(c_3 r) 
%\exp \left(-\frac{r^2}{2R^2}\right)dr 
%\nonumber\\
=\frac{ C_N C_0}{\kappa T} && 2\pi e^{-p^0/T}
\int\limits^{aY}_{-aY} \exp\big(c_1 y-c_2 y^2\big) dy
\nonumber\\
&&\times\int\limits^{bR}_{0}  \ r\, I_0(c_3 r)\, \exp(-c_4 r^2) dr\;  ,
\label{e21}
\ea
where
$c_1=p_y \dot Y/(YT)$, 
$c_2=1/(2Y^2)$,  
$c_4=1/(2R^2)$ are constants.

\begin{figure}[h]  %%%%%%%%%%
\begin{center}
\resizebox{0.9\columnwidth}{!}
{\includegraphics{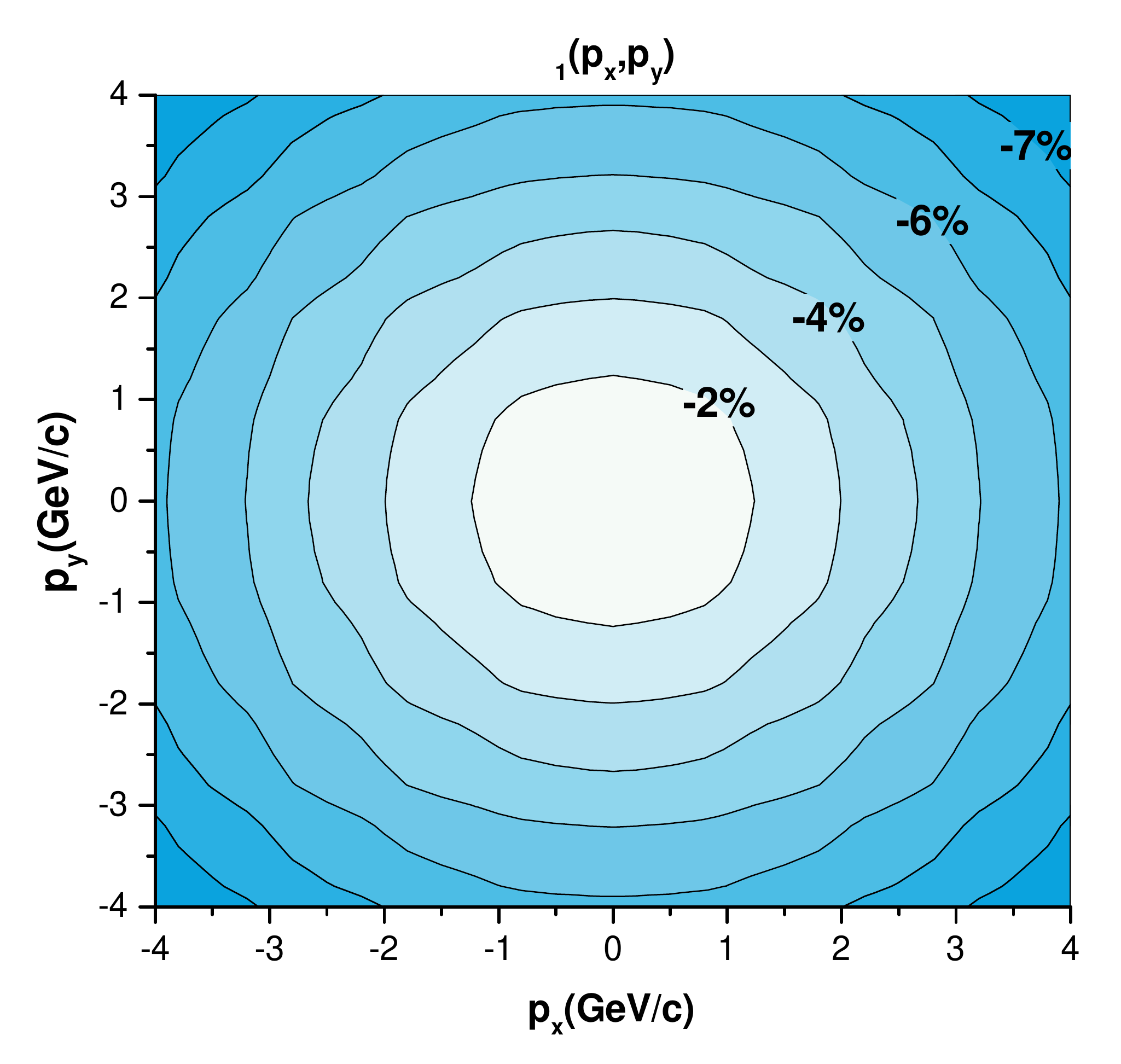}}
\caption{
(color online) 
The polarization of $\Lambda$ particles, $\vec \Pi_1(\vec p)$, 
in the participant Center of Mass (CM) frame for the
first term containing the $(\nabla \times \vec \beta)$-contribution,
at time $t =0.5$ fm/c after the equilibration of the rotation,
in the Exact model. The polarization, $\vec \Pi_1(\vec p)$,
points into the $-y$-direction and
changes from $-1.5$\% at the CM-momentum ($p_x=p_y=0$), to $-8$\% in the
corners, in $1$\% steps per contour line. The negative percentage indicates
that the polarization is in the $-y$-direction.
The structure is just like that of the energy weighted
vorticity. Due to azimuthal symmetry of the Exact Model the $p_x$ and $p_z$
dependence of $\vec \Pi$ are the same.
}
\label{F-P1_xy}
\end{center}
\end{figure}
%%%%%%%%%%%%%%%%%%%%%%%%%%%%%%%%%%%
%
Now we assume an infinite system 
with scaling Gaussian density profile, 
so that the integrals are evaluated up to infinity, 
i.e. the parameters $a=\infty$, $b=\infty$.
Thus, the $y$ component integration in Eq. (\ref{e21}) is calculated as:
\ba
&&\int^{+\infty}_{-\infty}e^{c_1y-c_2y^2}dy
%\nonumber\\
%&=&\frac12 \sqrt{\frac{\pi}{c_2}}
%\exp\Big(\frac{c_1^2}{4c_2^2}\Big)
%\bigg[{\rm erf}\Big(\sqrt{c_2}y -
%\frac{c_1}{2\sqrt{a}}\Big)\bigg]^{+\infty}_{-\infty}
%\nonumber\\
=\sqrt{\frac{\pi}{c_2}}
\exp\Big(\frac{c_1^2}{4c_2^2}\Big),
\label{e21a}
\ea
where we used  the integral formula No. 2.33(1) in \cite{AlZ07}, 
and $\rm erf \, (+\infty)=1$, $\rm erf  \,(-\infty)=-1$.

For the integration of $r$ component:
\ba
&&\int^{+\infty}_{0} r \, I_0(c_3 r)\,e^{-c_4 r^2} dr
\nonumber\\
&=&\frac{1}{c_3\sqrt{c_4}}
\exp\Big(\frac{c_3^2}{8c_4}\Big) \ 
M_{-\frac12,0}\Big(\frac{c_3^2}{4c_4}\Big),
\label{e22}
\ea
where the $M_{-\mu,\nu}(z)$ is the so called 'Whittaker Function', 
 No.6.643(2) in \cite{AlZ07}.

Now, we obtain the final form of Eq. (\ref{e21a}):
\ba
A(p)
% &=& \int dV\ n_F(p,s)
%\nonumber\\
%&=&\frac{ C_N C_0}{\kappa T}\, 2\pi e^{-p^0/T}
%\times
%\sqrt{\frac{\pi}{c_2}}
%\exp\Big(\frac{c_1^2}{4c_2^2}\Big)
%\nonumber\\
%\quad&&\times\,
%\frac{1}{c_3\sqrt{c_4}}
%\exp\Big(\frac{c_3^2}{4c_4}\Big) \ 
%M_{-\frac12,0}\Big(\frac{c_3^2}{8c_4}\Big)
%\nonumber\\
&=&\frac{2\pi \sqrt{\pi}}{\kappa T}\frac{C_N C_0}{c_3\sqrt{c_2 c_4}}
e^{-p^0/T}
\exp\Big(\frac{c_1^2}{4c_2^2}\Big)
\nonumber\\
\quad&&\times
\exp\Big(\frac{c_3^2}{8c_4}\Big) \ 
M_{-\frac12,0}\Big(\frac{c_3^2}{4c_4}\Big).
\label{e23}
\ea

However, in the relativistic case, the integrations 
with respect to  $y$ and $r$ can not be performed
analytically, because of the presence of the factor
$\gamma=1/\sqrt{1-v^2_r- v^2_y-v^2_\phi}$\ \ .

%%Compared with denominator, the numerator of $\Pi(\vec{p})$  
%%has an additional term 
%%$(\bigtriangledown \times \vec{\beta})$ 
%%in the integral, there is no reason that it has an analytic solution. 
%%{\color{red}
%%Therefore, as a conclusion, the analytic solution is not feasible.
%%} % End color red

%%%%%%%%%%%%%%%%%%%%%%%%%%%%%%%%%%%
\subsection{The numerator}
%%%%%%%%%%%%%%%%%%%%%%%%%%%%%%%%%%%

Ref. \cite{CsI15} calculates the energy weighted 
vorticity, which is azimuthally symmetric, i.e. independent 
of the azimuthal angle $\phi$. 
In the definition of the polarization, Eq. (\ref{Pol-1}), we have 
$ p^0\,n_F(p,x) = \epsilon\, n_F(p,x)$ for $\Lambda$s with momentum $p$. 
In \cite{CsI15}, however,
the energy weighting is performed with the total energy density
of the fluid $E_{tot} = E_{int} + E_{kin}$, which in general is not 
the same as  $\epsilon\,n_F(p,x)$. On the other hand the bare vorticity
is just a constant in the non-relativistic Exact model, while the
EoS may be more general and it may  lead to more involved $R(t)$ and $Y(t)$
dependence than the ideal J\"uttner gas approximation would allow.

\begin{figure}[h]  %%%%%%%%%%
\begin{center}
\resizebox{0.9\columnwidth}{!}
{\includegraphics{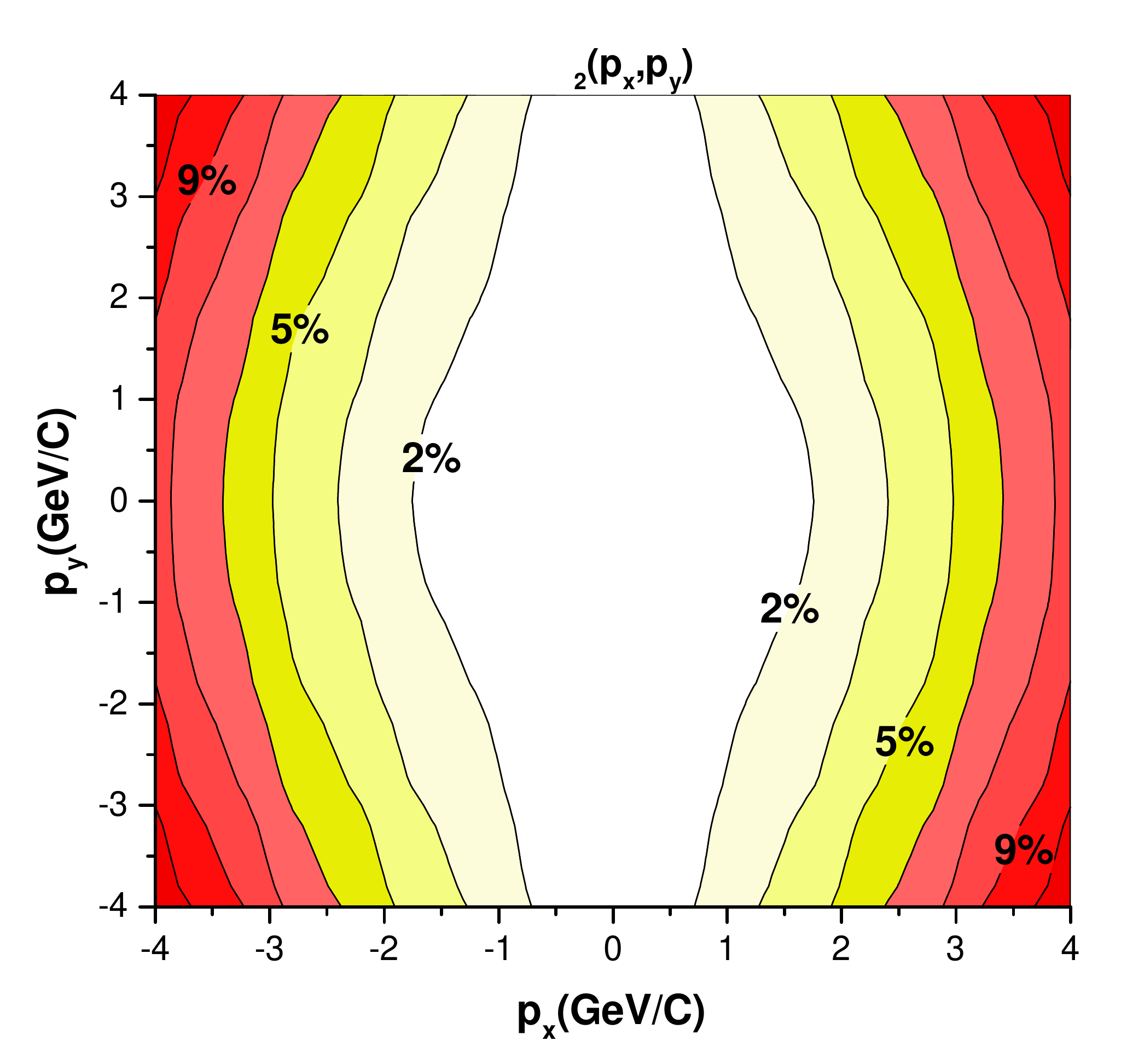}}
\caption{
(color online) 
The absolute value of $\Lambda$-polarization, $\vec \Pi_2(\vec p)$, 
in the participant Center of Mass (CM) frame for the
second term containing the $(\partial_t \vec \beta)$-contribution,
at time $t =0.5$ fm/c after the equilibration of the rotation,
in the Exact model. The polarization 
changes from zero at the CM-momentum ($p_x=p_y=0$), up to $20$\% in the
corners at $p_x=-4$GeV/c, in $2.5$\% steps per contour line.
In the corners at $p_x=4$GeV/c, the polarization is $12$\%.
This second term is orthogonal to  $\vec p$, 
and it is smaller, especially at CM-momenta, where it is negligible. 
This term arises from the expansion, which is
increasing rapidly in the Exact model with time and also increases
with the radius. At large radius the larger expansion leads to
larger momenta.
The structure of the 2nd component of polarization arises 
from the asymmetries of the different components of $\vec \Pi_2(\vec p)$\\
%{\color{red} }
}
\label{F-P2_xy}
\end{center}
\end{figure}
%%%%%%%%%%%%%%%%%%%%%%%%%%%%%%%%%%%

Thus we use the direct, non-relativistic 
vorticity values, $\omega(t)$, from Ref. \cite{CsI15},
and not the presented energy weighted vorticity. I.e. 
\be
\vec \nabla \times \vec \beta  = - 2\, \omega(t)\, \vec e_y / T(t) \ ,
\ee
so that the thermal vorticity has only $y$-directed component in the
Exact model. With the model parameters mentioned above 
(beginning of sec. \ref{S-II}), the thermal vorticity is 
$\hbar (\vec \nabla \times \vec \beta) = - 0.13$ at $t = 0.5$ fm/c, and 
it decreases very slowly with time, about 1-2\% per 1 fm/c. 
This constant vorticity will make the numerator simple:
\be
B(p) \equiv  \int \,d V \, n_F \big( \nabla \times \beta \big)
%&=& \int\limits_0^R r \,d r \int\limits_{-Y}^{+Y} \,d y
% \int\limits_0^{2\pi} \,d \phi\ n_F(x,p)
% \big( \frac{ - 2 \omega \vec e_y}{T} \big) \ .
%\nonumber\\
%&=&\frac{ - 2 \omega \vec e_y}{T}
% \int\limits_0^R r \,d r \int\limits_{-Y}^{+Y} \,d y
% \int\limits_0^{2\pi} \,d \phi\ n_F(x,p)
%\nonumber\\
 = \frac{ - 2 \omega \vec e_y}{T}
\times A(p)
\ee

Therefore, the first term of polarization vector, 
i.e. Eq. (\ref{Pol-1}) will be:
\be
\vec{\Pi}_1(p) =
- \frac{\hbar\epsilon}{8m}
\frac{\int \,d V n_F(x,p)\, (\nabla {\times}\vec{\beta})}
{\int \,d V n_F(x,p) } =
\frac{\hbar\epsilon \omega}{4mT}\vec e_y ,
\label{e25}
\ee
which means the polarization vector  arising from the vorticity,
$\vec{\Pi}_1 (p)$,
in the Exact rotation model is a 
constant, (although time dependent), and parallel to the $y$-axis.

\begin{figure}[h]  %%%%%%%%%%
\begin{center}
\resizebox{0.9\columnwidth}{!}
{\includegraphics{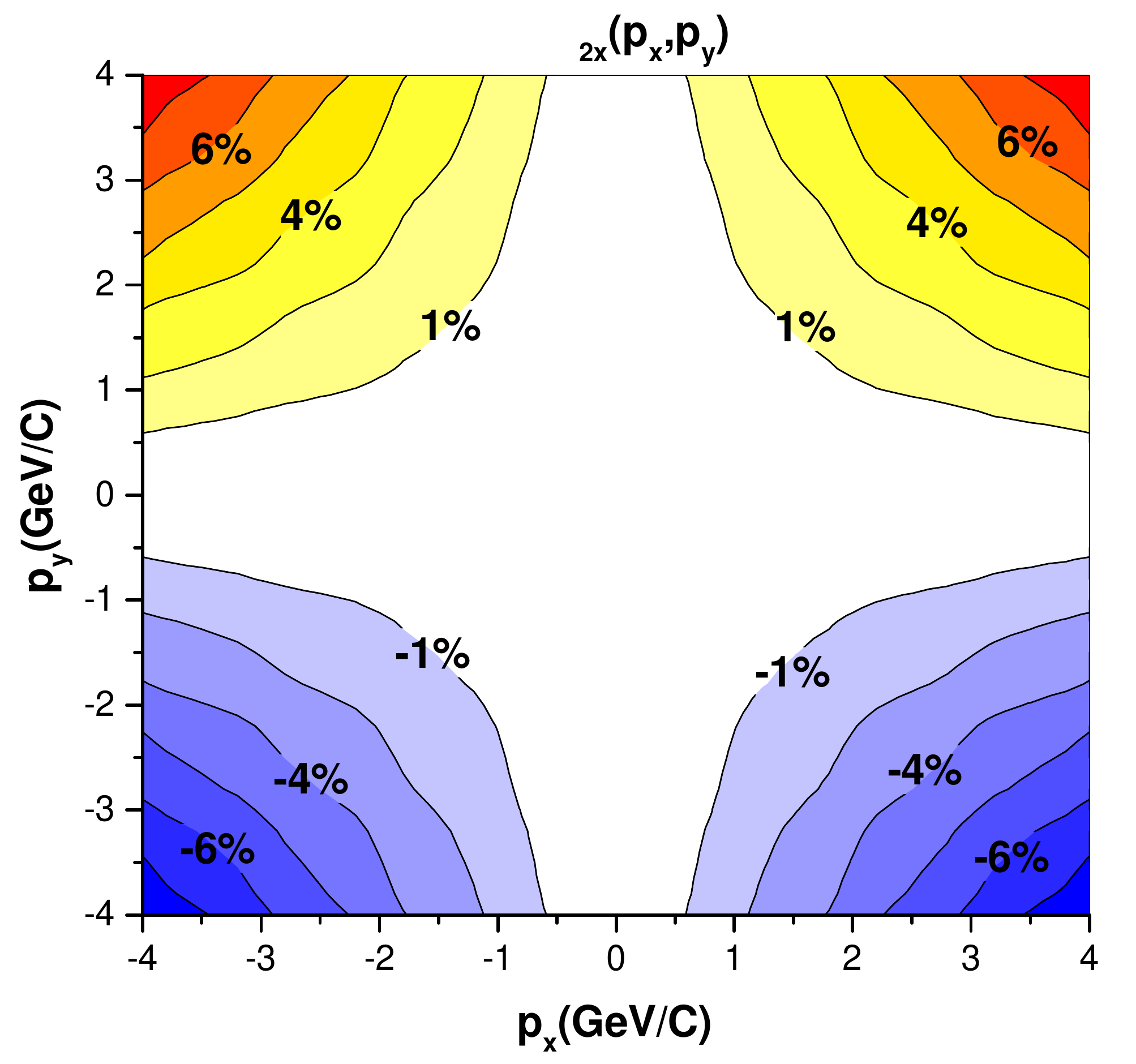}}
\caption{
(color online) 
The $x$ component of the $\Lambda$-polarization, $\vec \Pi_{2x}(\vec p)$, 
in the participant Center of Mass (CM) frame for the
second term containing the $(\partial_t \vec \beta)$-contribution,
at time $t =0.5$ fm/c after the equilibration of the rotation,
in the Exact model. The polarization vanishes at the CM-momentum 
($p_x=p_y=0$), and changes from zero  up/down to $\pm 8$\% in the
corners, in $1$\% steps per contour line.
This term arises from the expansion, which is
increasing rapidly in the Exact model with time and also increases
with the radius. At large radius the larger expansion leads to
larger momenta.}
\label{F-P2x_xy}
\end{center}
\end{figure}
%%%%%%%%%%%%%%%%%%%%%%%%%%%%%%%%%%%

One may add the Freeze-Out (FO) probability to the integral.
According to the Ref. \cite{CsVW14}, the FO probability is
$
w_s = (p_\mu\,\hat\sigma^\mu_s)  ~  ( \vec p \cdot \vec u(x) ) ,
$ 
where the approximation is used that the FO direction, 
$\hat\sigma^\mu_s$ is parallel to the flow velocity,  
$\vec u(x) = \gamma \vec v(x)$.
In the first term of the numerator, which depends on the
constant $y$-directed vorticity this FO probability influences
the numerator and denominator the same way, so the effect of the 
two integrals cancel each other in the FO probability also.

%%%%%%%%%%%%%%%%%%%%%%%%%%%%%%%%%%%
\subsection{The second term}
%%%%%%%%%%%%%%%%%%%%%%%%%%%%%%%%%%%

The numerator in second term of polarization vector reads:
\be
\vec C(\vec p) \equiv \int \,d V n_F(x,p)\, 
(\partial_t \vec{\beta} + \vec \nabla \beta^0) \ .
\label{Pol3}
\ee
If, in the non-relativistic limit,
$\gamma = 1$ is assumed, then $\nabla \beta^0= 0$, 
and $\partial_t \beta = \partial_t (\vec{v} / T)$,
so we have to evaluate only the first term of the sum in the integrand.
According to Ref. \cite{CsI15,Stoecker_handbook}, 
the time derivatives of velocity are:
\ba
\partial_t v_r &=&
\bigg[ 
\big(\frac{\ddot{R}}{R} - \frac{\dot{R}^2}{R^2} \big)
- \omega^2  \bigg] r \equiv  c_5 r
\nonumber\\
\partial_t v_\phi &=&
\left(\dot{\omega} +  2 \frac{\dot{R}}{R}\omega \right) r \equiv  c_6 r
\nonumber\\
\partial_t v_y &=&
\bigg[ 
\frac{\ddot{Y}}{Y} - \frac{\dot{Y}^2}{Y^2} \bigg] y
 \equiv  c_7 y \ ,
\ea
where
 $c_5=(\ddot{R}/R - \dot{R}^2 /R^2 - \omega^2)$,
 $c_6 =(\dot{\omega} +  2 (\dot{R}/R)\omega )$, and
 $c_7=(\ddot{Y}/Y - \dot{Y}^2 /Y^2)$. 
\begin{figure}[h]  %%%%%%%%%%
\begin{center}
\resizebox{0.9\columnwidth}{!}
{\includegraphics{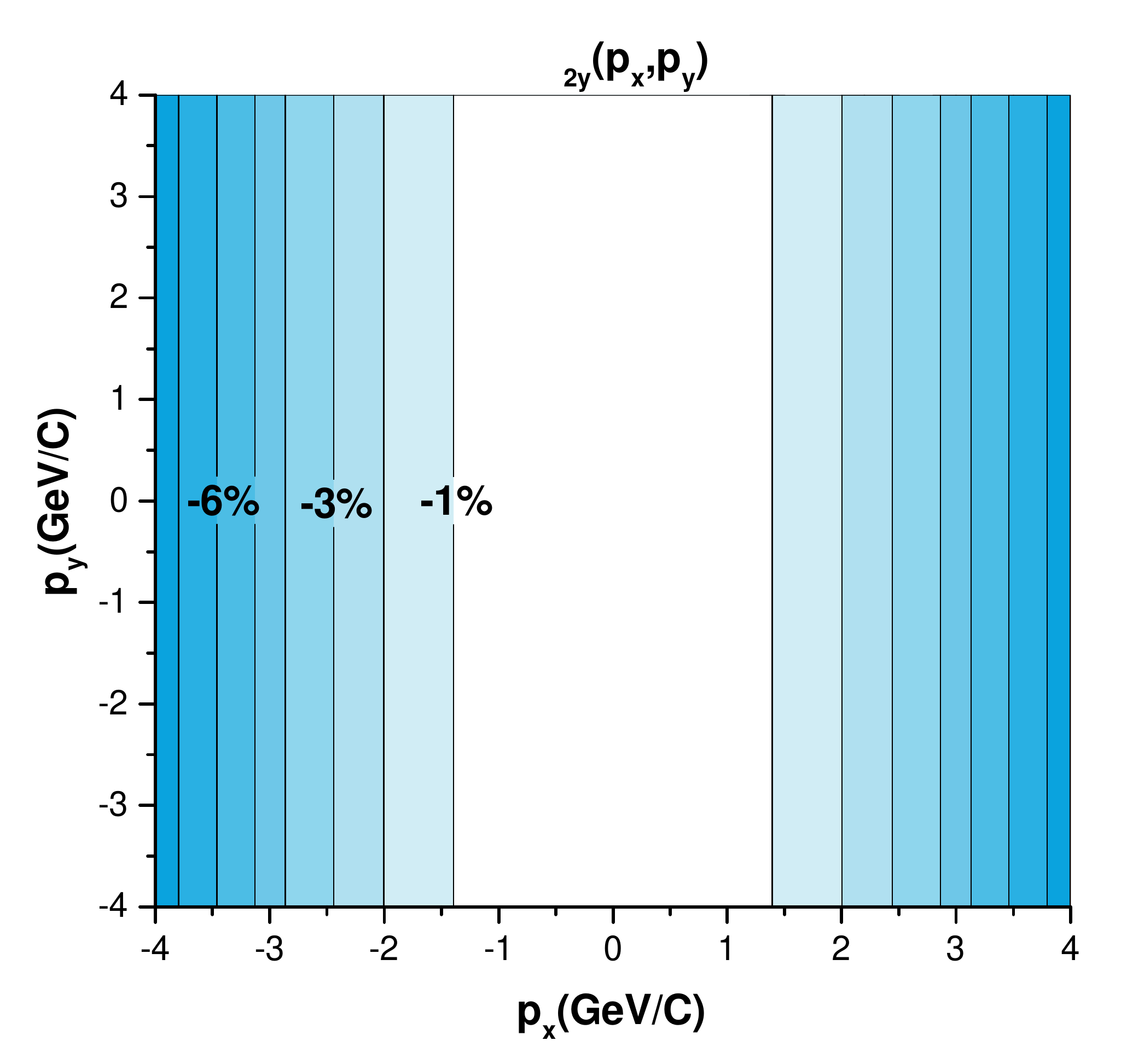}}
\caption{
(color online)
The $y$-component of $\Lambda$-polarization, $\vec \Pi_2(\vec p)$, 
in the participant Center of Mass (CM) frame for the
first term containing the $(\partial_t \vec \beta)$-contribution,
at time $t =0.5$ fm/c after the equilibration of the rotation,
in the Exact model. 
The polarization changes from zero in the middle to $-8$\% 
at $p_x=\pm 4$GeV/c, in $1$\% steps per contour line.
This $y$-component points into the axis-direction just as the first term,
$\vec \Pi_1$, thus these two are additive. 
The $y$-component of $\vec \Pi_2(\vec p)$ does not depend of $p_y$, as 
shown in Eq. (\ref{e38}).}
\label{F-P2y_xy}
\end{center}
\end{figure}
%%%%%%%%%%%%%%%%%%%%%%%%%%%%%%%%%%%

Therefore, 
$$
\partial_t \vec \beta =
\big( c_5 r\vec{e}_r 
+ c_6 r \vec{e}_\phi 
+ c_7 y \vec{e}_y \big) /T \ ,
$$
provides the time-components of the vorticity in the three spatial directions.
Here, as the model is symmetric, $\partial_t \beta_y$ vanishes, and
with the model parameters mentioned above (sec. \ref{S-II}),
at $t = 0.5$ fm/c and $r = 1$ fm
$ \frac{\hbar}{c} \partial_t \beta_r = 0.024$ and
$ \frac{\hbar}{c} \partial_t \beta_\phi = 0.009$. Both these vorticity
components decrease slowly with time by about $0.0005$ in 1 fm/c.

Eq. (\ref{Pol3}) is a volume integral of a vectorial quantity, 
which is not convenient to perform in cylindrical coordinates.
So we transform it into Cartesian coordinates: 
$\vec{e}_r = \cos \phi\, \vec{e}_x + \sin \phi\, \vec{e}_z$, 
$\vec{e}_\phi = -\sin \phi\, \vec{e}_x + \cos \phi\, \vec{e}_z$. 
Therefore, 
$
 T \cdot \partial_t \vec{\beta}
=
\big( c_5 \cos \phi - c_6 \sin \phi \big) r \, \vec{e}_x
 + 
\big( c_5 \sin \phi + c_6 \cos \phi \big) r \, \vec{e}_z
 +  c_7 y  \, \vec{e}_y \ .
$

The integral of Eq. (\ref{Pol3}) can be expanded as:
\ba
&&\vec C(\vec{p}) = \int \,d V n_F(x,p)\, \partial_t \vec{\beta}
\nonumber\\
%% &&=\frac{C_N C_0 e^{-p_0/T}}{\kappa T}
%%   \int^{aY}_{-aY}\!\!\!\!\!\!   dy\,
%%    \int^{bR}_{0}\!\!\!\!\!\!\! r dr\, 
%\exp\left(-\frac{y^2}{2Y^2}\right) \exp\left(-\frac{r^2}{2R^2}\right) 
%% \nonumber\\
%% &&\times \int^{2\pi}_0 \!\!\!\!\!\!   d\phi\,  
%%    \exp\left(\frac{p_x\dot{R}}{TR}r \cos \phi -
%%    \frac{p_x\omega }{T}r \sin \phi + 
%%    \frac{r_y \dot{Y}}{TY}y\right)
%%    \partial_t \vec{\beta}
%% \nonumber\\
&&=\frac{C_N C_0}{\kappa T}  e^{-p_0/T}
   \iiint rdr d\phi dy\,
   \exp \left( c_1 y - c_2 y^2 \right)
\nonumber\\
&&\qquad\qquad    \times 
   \exp \left( a \cos \phi - b \sin \phi - c_4 r^2 \right)
   \partial_t \vec{\beta} \ ,
\label{e28}
\ea
where $a$ and $b$ are defined after Eq. (\ref{eab}).

\begin{figure}[ht]  %%%%%%%%%%
\begin{center}
\resizebox{0.9\columnwidth}{!}
{\includegraphics{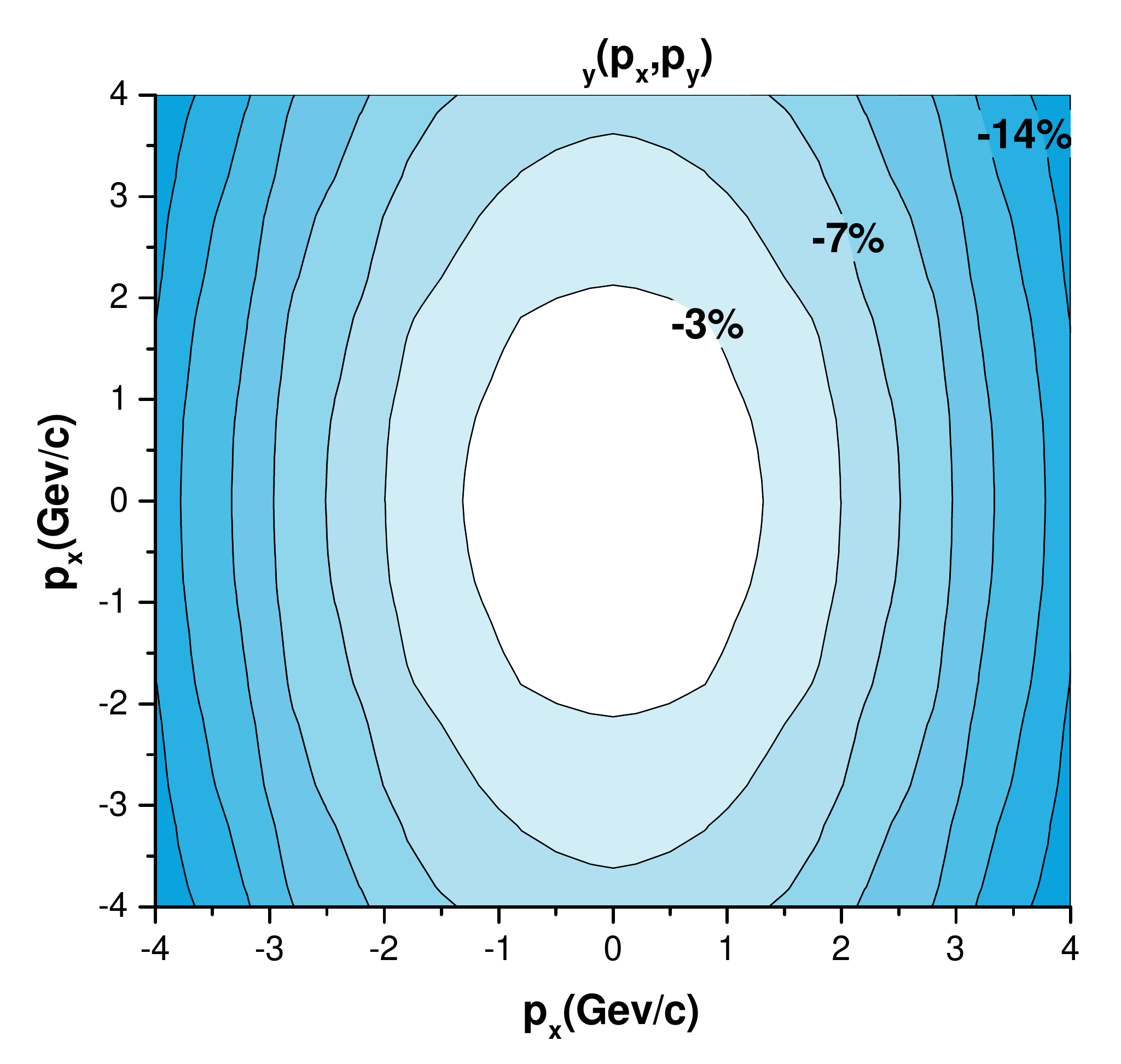}}
\caption{
(color online)
The $y$-component of $\Lambda$-polarization, $\vec \Pi(\vec p)$, 
in the participant Center of Mass (CM) frame for the
second term containing the $(\partial_t \vec \beta)$-contribution,
at time $t =0.5$ fm/c after the equilibration of the rotation,
in the Exact model. 
The polarization is -1.5\% at the CM-momentum ($p_x=p_y=0$), 
it is  $-16$\% in the corners.
The change is in steps of  $2$\% per contour line.}
\label{F-Py_xy}
\end{center}
\end{figure}
%%%%%%%%%%%%%%%%%%%%%%%%%%%%%%%%%%%

It is convenient to define an integrating operator, 
$\bar{A}$ as:
\ba
\bar{A}&=& \int \,d V n_F(x,p)\, \times
\nonumber\\
 &=&  \iiint r\,dr\, d\phi\, dy\, e^{c_1 y - c_2 y^2} 
      e^{a \cos \phi - b \cos \phi - c_4 r^2}\times
\ ,
\nonumber
\ea
and then Eq. (\ref{e28}) will be:
\be
\vec C(\vec{p}) = \bar{A}\cdot \partial_t \vec{\beta}
%\nonumber\\  =
%\bar{A} \cdot \frac 1T
%\big( c_5 \cos \phi - c_6 \sin \phi \big) r \, \vec{e}_x
%\nonumber\\
%&+& \bar{A} \cdot \frac 1T
%\big( c_5 \sin \phi + c_6 \cos \phi \big) r \, \vec{e}_z
%\nonumber\\
%&+& \bar{A} \cdot \frac 1T c_7 y  \, \vec{e}_y 
%\nonumber\\
 \equiv  \frac 1T \left( I \vec{e}_x + J  \vec{e}_z + H\vec{e}_y \right)\ ,
\label{e29}
\ee
where we defined:
\ba
I &\equiv& \bar{A} \cdot \big( c_5 \cos \phi - c_6 \sin \phi \big) r \ ,
\nonumber\\
J &\equiv& \bar{A} \cdot \big( c_5 \sin \phi + c_6 \cos \phi \big) r \ ,
\nonumber\\
H &\equiv& \bar{A} \cdot  c_7 y \ . \nonumber
\ea

% $H$ can be expanded as:
% \ba
% H&=&\bar{A} \cdot  c_7 y
% \nonumber\\
% &=&\frac{C_N C_0}{\kappa T} e^{-p_0/T} 
%  \int dy \,
%    c_7\, y \exp \big( c_1 y -c_2 y \big)
% \nonumber\\
% && \times
% \iint rdr \, d\phi \, 
% \exp \Big[ (a \cos \phi - b \sin \phi ) - c_4 r^2 \Big] \ .
% \nonumber\\
% \label{e30}
% \ea

% According to Eqs. (\ref{e18}) and (\ref{e22}), 
% the integral with respect to $r$ and $\phi$ is:
% \ba
% &&\iint rdr \, d\phi \, 
% \exp \Big[ (a \cos \phi - b \sin \phi ) - c_4 r^2 \Big]
% \nonumber\\
% =&&\frac{2\pi}{c_3\sqrt{c_4}}
% \exp\Big(\frac{c_3^2}{8c_4}\Big)  \ 
% M_{-\frac12,0}\Big(\frac{c_3^2}{4c_4}\Big) \ ,
% \nonumber
% \ea
% and the integral with respect to $y$ is calculated as:
% \ba
% \int dy \,  c_7 \, y \exp \left( c_1 y -c_2 y^2 \right)
% = \frac{c_7 c_2}{2c_2}
%   \sqrt{\frac{\pi}{c_2}}
%   \exp\Big(\frac{c_1^2}{4c_2^2}\Big) \ ,
% \nonumber
% \ea
 Using the integral formula No. 2.33(6) of 
\cite{AlZ07} the function $H$ becomes:
\ba
H=&&\frac{2\pi\sqrt{\pi}C_N C_0}{\kappa T} e^{-p_0/T} 
  \frac{c_7 c_1}{2c_3 c_2 \sqrt{c_4 c_2}}
  \nonumber\\
&&\times  \exp\Big(\frac{c_3^2}{8c_4}\Big)
  \exp\Big(\frac{c_1^2}{4c_2^2}\Big)
  M_{-\frac12,0}\Big(\frac{c_3^2}{4c_4}\Big) \ .
\label{e31}
\ea

\bigskip
The function 
$I$ can be expanded as a function of integrals over
$\phi$, $r$ and $y$. The integral over $\phi$ brings in the
Bessel function, $2 \pi c_8 I_1(c_3 r)/c_3$ (See No. 3.937 (1) and (2)
of \cite{AlZ07}), where
$c_8=( c_5 a'  - c_6 b' )$, 
$a' =a/r=|p_x| \dot{R} /TR $, and 
$b'  =b/r=|p_x| \omega /T $.
Subsequently, the integral with respect to $r$ brings in the  
'Whittaker Function' and then the final form of $I$ 
%
% \ba
% &&I=\bar{A}\cdot \bigg(c_5 \cos \phi - c_6 \sin \phi \bigg) r
% \nonumber\\
% &&=\frac{C_N C_0}{\kappa T} e^{-p_0/T} 
% \int dy \exp \Big( c_1 y - c_2 y^2 \Big) \,
% \iint r dr \, d\phi \, 
% \nonumber\\
% &&\times \exp \Big[ (a \cos \phi - b \sin \phi ) - c_4 r^2 \Big]
% \bigg(c_5 \cos \phi - c_6 \sin \phi \bigg) r
% \nonumber\\
% &&=\frac{C_N C_0}{\kappa T} e^{-p_0/T} 
% \int dy \exp \big( c_1 y - c_2 y^2 \big)
% \int dr \, r^2 e^{-c_4 r^2}
% \nonumber\\
% && \quad\times \int  d\phi
% \exp \bigg[ a \cos \phi - b \sin \phi\bigg]
% \bigg(c_5 \cos \phi - c_6 \sin \phi \bigg) \ ,
% \nonumber\\
% \label{e32}
% \ea
% According to Eqs. No. 3.937 (1) and (2) in Ref. \cite{AlZ07}, 
% one can first perform the integration with respect to $\phi$:
% \ba
% &&\int^{2\pi}_{0}  d\phi \exp \bigg[ a \cos \phi - b \sin \phi\bigg]
% \bigg(c_5 \cos \phi - c_6 \sin \phi \bigg)
% \nonumber\\
% =&& \quad \frac{c_8}{c_3} 2\pi I_1(c_3 r) \ , 
% \nonumber
% \ea
% where $c_8=( c_5 a'  - c_6 b' )$, 
% and $a' =a/r=|p_x| \dot{R} /TR $, 
% $b'  =b/r=|p_x| \omega /T $.
% Then, the integral with respect to $r$ becomes:
% \ba
% &&\int^{\infty}_{0} dr r^2 e^{-c_4 r^2} \cdot \frac{c_8}{c_3} 2\pi I_1(c_3 r)
% \nonumber\\
% &=& 2\pi \frac{c_8}{c_3}
%   \int^{\infty}_{0} dr r^2 I_1(c_3 r) e^{-c_4 r^2}
% \nonumber\\
% &=&2\pi \frac{c_8}{c_3^2 c_4}
% \exp \bigg( \frac{c_3^2}{8c_4} \bigg) \,
% M_{-1,\frac 12}\Big( \frac{c_3^2}{4c_4} \Big)\ ,
% \label{e33}
% \ea
% where we used the 6.643(2) of Ref. \cite{AlZ07}.
% 
after performing the separable integration with respect to $y$ 
leads to:
\ba
I=&&\frac{2\pi\sqrt{\pi}C_N C_0}{\kappa T} e^{-p_0/T} 
  \frac{c_8}{c_3^2 c_4 \sqrt{c_2}}
  \nonumber\\
&&\times \exp\Big(\frac{c_3^2}{8c_4}\Big)
         \exp\Big(\frac{c_1^2}{4c_2^2}\Big)\,
          M_{-1,\frac 12}\Big( \frac{c_3^2}{4c_4} \Big)\ .
\label{e33}
\ea

\bigskip
Evaluating the  integral $J$ is similar to $I$:
\ba
J=&&\frac{2\pi\sqrt{\pi}C_N C_0}{\kappa T} e^{-p_0/T} 
  \frac{c_9}{c_3^2 c_4 \sqrt{c_2}}
  \nonumber\\
&&\times \exp\Big(\frac{c_3^2}{8c_4}\Big)
         \exp\Big(\frac{c_1^2}{4c_2^2}\Big)
          M_{-1,\frac 12}\Big( \frac{c_3^2}{4c_4} \Big)\ .
\label{e34}
\ea
where the only difference is: $c_9=( c_5 b'+ c_6 a')$ 
compared to $c_8$ in $I$.

Then, substituting $I$, $J$, $H$ back into Eq. (\ref{e29}), 
one can obtain the analytical solution for numerator in 
second term of polarization vector as: 
\ba
&&\vec C(\vec{p})= \int \,d V n_F(x,p)\, \partial_t \vec{\beta}
= \frac 1T \left( I \vec{e}_x {+} J  \vec{e}_z {+} H\vec{e}_y \right)
\nonumber\\
&=&\frac{2\pi\sqrt{\pi}C_N C_0}{\kappa T^2} e^{-p_0/T} 
   \exp\Big(\frac{c_3^2}{8c_4}\Big)
   \exp\Big(\frac{c_1^2}{4c_2^2}\Big) \times
  \nonumber\\
&& \Bigg[
         \frac{c_8}{c_3^2 c_4 \sqrt{c_2}}
         M_{-1,\frac 12}\big( \frac{c_3^2}{4c_4} \big)
         \vec{e}_x +
         \frac{c_9}{c_3^2 c_4 \sqrt{c_2}}\,
         M_{-1,\frac 12}\Big( \frac{c_3^2}{4c_4} \Big) \vec{e}_z
\nonumber\\
&&\qquad\qquad\quad\,\,
+ \frac{c_7 c_1}{2c_3 c_2 \sqrt{c_4 c_2}}\,
  M_{-\frac12,0}\Big(\frac{c_3^2}{4c_4}\Big) \vec{e}_y
\Bigg] \ .
\nonumber\\
\ea
Dividing this by $A(\vec{p})$, i.e. Eq. (\ref{e23}), one gets:
\ba
 \frac{\vec C(\vec{p})}{A(\vec{p})}
 {=}\frac{1}{T} \Big[
\frac{c_8}{c_3 \sqrt{c_4}}
\frac{M_{-1,\frac 12}}{M_{-\frac 12,0}}
\vec{e}_x {+} 
\frac{c_9}{c_3 \sqrt{c_4}}
\frac{M_{-1,\frac 12}}{M_{-\frac 12,0}}
\vec{e}_z {+}
\frac{c_7 c_1}{2c_2}\vec{e}_y
\Big].
\nonumber\\
\ea

\begin{figure}[ht]  %%%%%%%%%%
\begin{center}
\resizebox{0.9\columnwidth}{!}
{\includegraphics{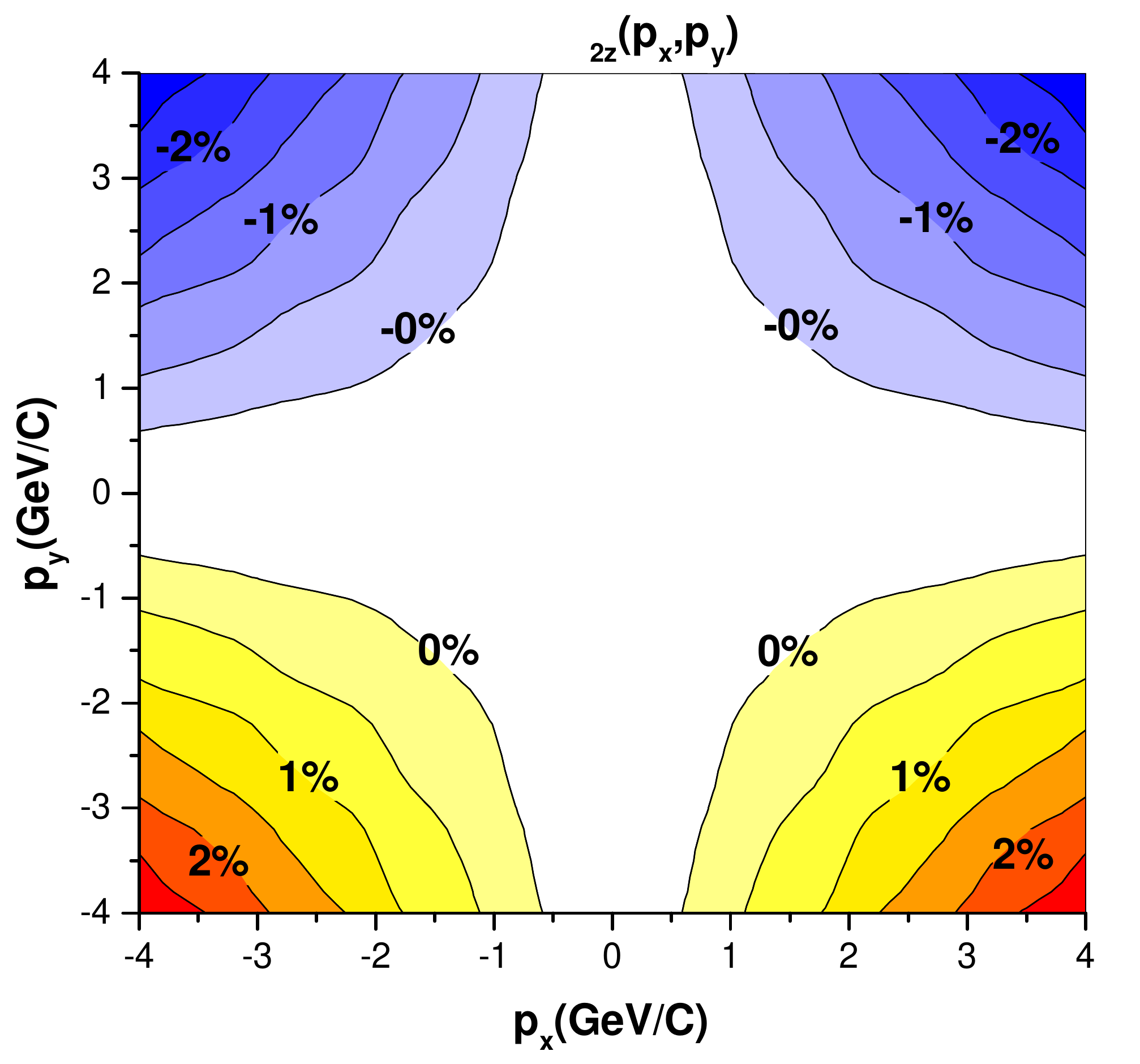}}
\caption{
(color online)
The $z$-component of $\Lambda$-polarization, $\vec \Pi_2(\vec p)$, 
in the participant Center of Mass (CM) frame for the
second term containing the $(\partial_t \vec \beta)$-contribution,
at time $t =0.5$ fm/c after the equilibration of the rotation,
in the Exact model. 
The polarization vanishes at the CM-momentum ($p_x=p_y=0$), 
it is  $\pm 3$\% in the corners.
The change is in steps of  $0.5$\% per contour line.
The corners at $p_y=-4$GeV/c are positive 
while at $p_y=4$GeV/c are negative. \\
}
\label{F-P2z_xy}
\end{center}
\end{figure}
%%%%%%%%%%%%%%%%%%%%%%%%%%%%%%%%%%%

Then, we obtain the second term of polarization vector:
\ba
\vec \Pi_2(\vec{p})
=&&\frac{\hbar \vec{p}}{8m}\times \frac{\vec C(\vec{p})}{A(\vec{p})}
\nonumber\\
=&&\frac{\hbar}{8mT}\Bigg[
\frac{p_y c_9}{c_3\sqrt{c_4}} 
\frac{M_{-1,\frac 12}}{M_{-\frac 12,0}}
\vec{e}_x -
\frac{|p_x| c_9}{c_3\sqrt{c_4}} 
\frac{M_{-1,\frac 12}}{M_{-\frac 12,0}}
\vec{e}_y 
\nonumber\\
&&\qquad\qquad \quad + \quad \Big(
\frac{|p_x| c_7 c_1}{2c_2} - 
\frac{p_y c_8}{c_3\sqrt{c_4}} 
\frac{M_{-1,\frac 12}}{M_{-\frac 12,0}} \Big)
\vec{e}_z 
\Bigg] .
\nonumber\\
\label{e38}
\ea

As we can see, and as is given also by the definition, Eq. (\ref{Pol-1}),
the second term of polarization is orthogonal to the particle momentum:
\be
\vec \Pi_2(\vec{p}) \ \perp \ \vec p \ ,
\ee
thus if we use the choice that $\vec p$ should be in the 
$[x,y]$-plane and its $z$-component should vanish, then the $y$-component
of $\vec \Pi_2(\vec{p})$, should depend on  $p_x$ only, 
see Fig. \ref{F-P2y_xy}.

%%%%%%%%%%%%%%%%%%%%%%%%%%%%%%%%%%%%%%%%%%%%%%%%%%%%%%%%%%%%%%%
\section{The freeze-out stage}
%%%%%%%%%%%%%%%%%%%%%%%%%%%%%%%%%%%%%%%%%%%%%%%%%%%%%%%%%%%%%%%

The fluid dynamical model is in principle not adequate to describe
the final, post freeze-out (FO) particle distributions, the
abundance of the particle species and also their polarization.
This is so because the post freeze-out distributions must not be
in local thermal equilibrium and must not have interactions 
among the final emitted particles.  Furthermore, the emitted
particles should not move back into the interacting zone, i.e. 
towards the pre-FO side of the FO hypersurface. 
How to handle the freeze-out is described in great detail
in \cite{ChengEtAl10}. It indicates two 
ways to handle this process: (i) consider the post-FO matter
as if it has an Equation of State (EoS). This is only possible if the
post FO EoS is that of a non-interacting ideal gas and the
FO hypersurface is timelike. (ii) The other approach is that 
the post FO matter is described by a dynamical model with
weak and rapidly decreasing interaction, like UrQMD or PACIAE,
matched to the QGP fluid on the FO hypersurface. The change at 
crossing this hypersurface is in general significant, as the 
pre-FO matter is strongly interacting, supercooled QGP, while the
post-FO matter is weakly interacting and has different (usually
less) degrees of freedom in both situations. The FO across  the
hypersurface is stronger if the latent heat of the transition is
larger. 

 The precise
way to perform this transition is described in \cite{ChengEtAl10}.
This method is demonstrated in several earlier fluid dynamical model
calculations (also using the PICR method), for precision calculations
of flow harmonics.

As mentioned in the introduction, at high energies
(RHIC and LHC) the Constituent 
Quark Number Scaling and the large strangeness abundance clearly indicate
a supercooling and rapid hadronization.  Furthermore at these energies
the transition is in the crossover domain of the EoS, thus the
expected changes are smaller, and the major part of the FO hypersurface
is time-like, which allows to use ideal gas post FO distributions,
as we do it here using the method of \cite{BeChZaGr13}. These 
are the conditions which make the changes in mechanical parameters
(e.g. $\vec v$) small at freeze-out while the temperature changes are
larger \cite{ChengEtAl10}.

Thus, just in the case of  Constituent Quark Number Scaling, we assume that 
other mechanical processes like mechanical polarization will not 
significantly change at freeze-out at RHIC and LHC energies.
This conclusion is restricted to local thermal and flow equilibrium,
and should not apply to some of the microscopic processes, which
dominate $p+p$ reactions.

Also, in case of freeze-out through space-like FO hypersurfaces,  the
mechanical parameters change significantly, the post-FO distribution
is far from a thermal distribution (it is a Cut-Juttner of Canceling
Juttner distribution), and thus the conditions of  \cite{BeChZaGr13}
 that we use, are not satisfied.

In this connection we may mention that in earlier related
publications, previous experimental $\Lambda$ polarization 
measurements, which were negative, were discussed. It was pointed out
that polarization as measured was averaged for all  $\Lambda$ particle
directions. Here, as well as in the
previous detailed PICR fluid dynamical calculations, it was emphasized
that polarization should be measured after finding Event by Event
the reaction plane and the center of mass of the system. 
Significant polarization can only expected for particles
emitted in selected directions.

Preliminary experimental polarization studies in the RHIC Beam Energy
Scan program along these lines are promising
\cite{Lisa},
and may lead soon to positive quantitative results. At this point of time
the present relatively simpler FO treatment of the model calculations with
constant time FO are sufficient, and can be refined when quantitative 
experimental data will be available.

%%%%%%%%%%%%%%%%%%%%%%%%%%%%%%%%%%%
\subsection{Conclusion}
%%%%%%%%%%%%%%%%%%%%%%%%%%%%%%%%%%%

Finally, adding Eqs. (\ref{e38}) and (\ref{e25})
we get the analytical solution for $\Lambda$-polarization in the Exact model:
\ba
\vec{\Pi}(p) &=&
\frac{\hbar}{8mT}\Bigg[
\frac{p_y c_9}{c_3\sqrt{c_4}} 
\frac{M_{-1,\frac 12}}{M_{-\frac 12,0}}
\vec{e}_x +
\Big( 2\epsilon \omega -
\frac{|p_x| c_9}{c_3\sqrt{c_4}} \times
\nonumber\\
&&
\frac{M_{-1,\frac 12}}{M_{-\frac 12,0}} \Big)
\vec{e}_y 
 + \Big(
\frac{|p_x| c_7 c_1}{2c_2} - 
\frac{p_y c_8}{c_3\sqrt{c_4}} 
\frac{M_{-1,\frac 12}}{M_{-\frac 12,0}} \Big)
\vec{e}_z 
\Bigg] \ .
\nonumber\\
\label{e39}
\ea

Notice that Eq. (\ref{e39}) is 
the analytical solution in the non-relativistic limit. 
The 'Whittaker Function', $M_{\mu,\nu}(z)$, is the 
confluent hypergeometric function.
For the relativistic case, the integrations 
of the $\Lambda$-polarization vector cannot be performed
analytically, because of the presence of 
$\gamma=1/\sqrt{1-v^2_r- v^2_y-v^2_\phi}$,
which will make the integrations more involved. 
Thus, a numerical solution for the $\Lambda$-polarization would be needed.

The effect of vorticity is shown in Fig. \ref{F-P1_xy}.  The non-relativistic
Exact model can handle reactions with modest energy and modest rotation, so
the overall vorticity and the resulting polarization is not too large.
Furthermore the rotation and vorticity decrease with time while the
radial and axial expansion increases. This expansion leads to the second
term of polarization, $\vec \Pi_2$, which depends on $\partial_t \vec \beta$
(while the $\vec \nabla \beta^0$ terms vanishes in the non-relativistic
approximation). Due to the simplicity of the Exact model, the vorticity
arising from the shear flow of the peripheral initial state is constant in 
space and depends on the time only. However, due to the construction of 
thermal vorticity, both the angular momentum and the temperature in the
denominator decrease with time, thus $\nabla \times \vec \beta$ is hardly
decreasing with the time, and it has a significant value, -0.13, in natural 
units. At the same time in this model the time-dependent vorticity is
smaller by almost an order of magnitude. The time-dependent vorticity 
components also decrease faster than the one originating from the initial
shear flow.

Nevertheless the second term in the polarization is of comparable 
magnitude to the term arising from local vorticity. See Fig. \ref{F-P2_xy}.

The presented plots are such that $p_x$ points into the direction of the
observed $\Lambda$-particle, while the $p_y$ is the axis direction.
All results should be either symmetric or antisymmetric for a $\pm p_y$
change.  On the other hand reversing the $p_x$ axis must not change the 
data, as the $x$-axis is chosen to be the direction of the argument of
$\vec \Pi(\vec p)$, which must be azimuthally symmetric in the $[x,y]$-plane.

The polarization arising from the dynamics of the 
radial and spherical expansion, $\vec \Pi_2$, was not discussed before
in the literature, as the dominance of the vorticity effect was anticipated
and studied up to now.
The $\vec \Pi_2$ plots in Figs. 
\ref{F-P2_xy}, \ref{F-P2x_xy}, \ref{F-P2y_xy}, \ref{F-P2z_xy},
show the components of the polarization arising from the dynamics of the 
spherical expansion. The most interesting $y$-component arises from the
$x$-component of the momentum and the $z$-component of the thermal
velocity change $\dot{\beta}_z$  (Fig. \ref{F-P2y_xy}).

\begin{widetext}

\begin{figure}[ht]  %%%%%%%%%%
\begin{center}
\resizebox{0.48\columnwidth}{!}
{\includegraphics{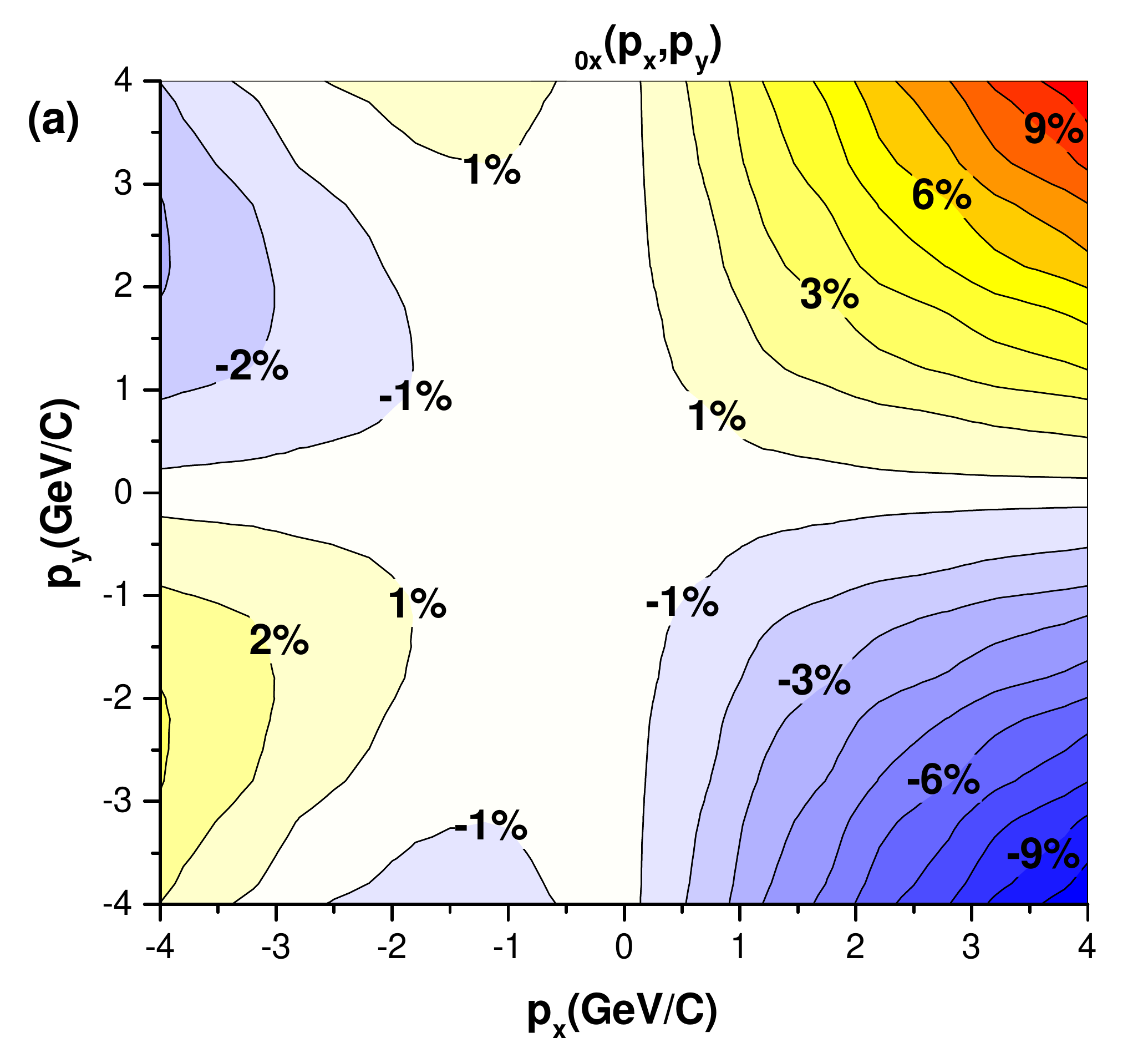}}\,
\resizebox{0.48\columnwidth}{!}
{\includegraphics{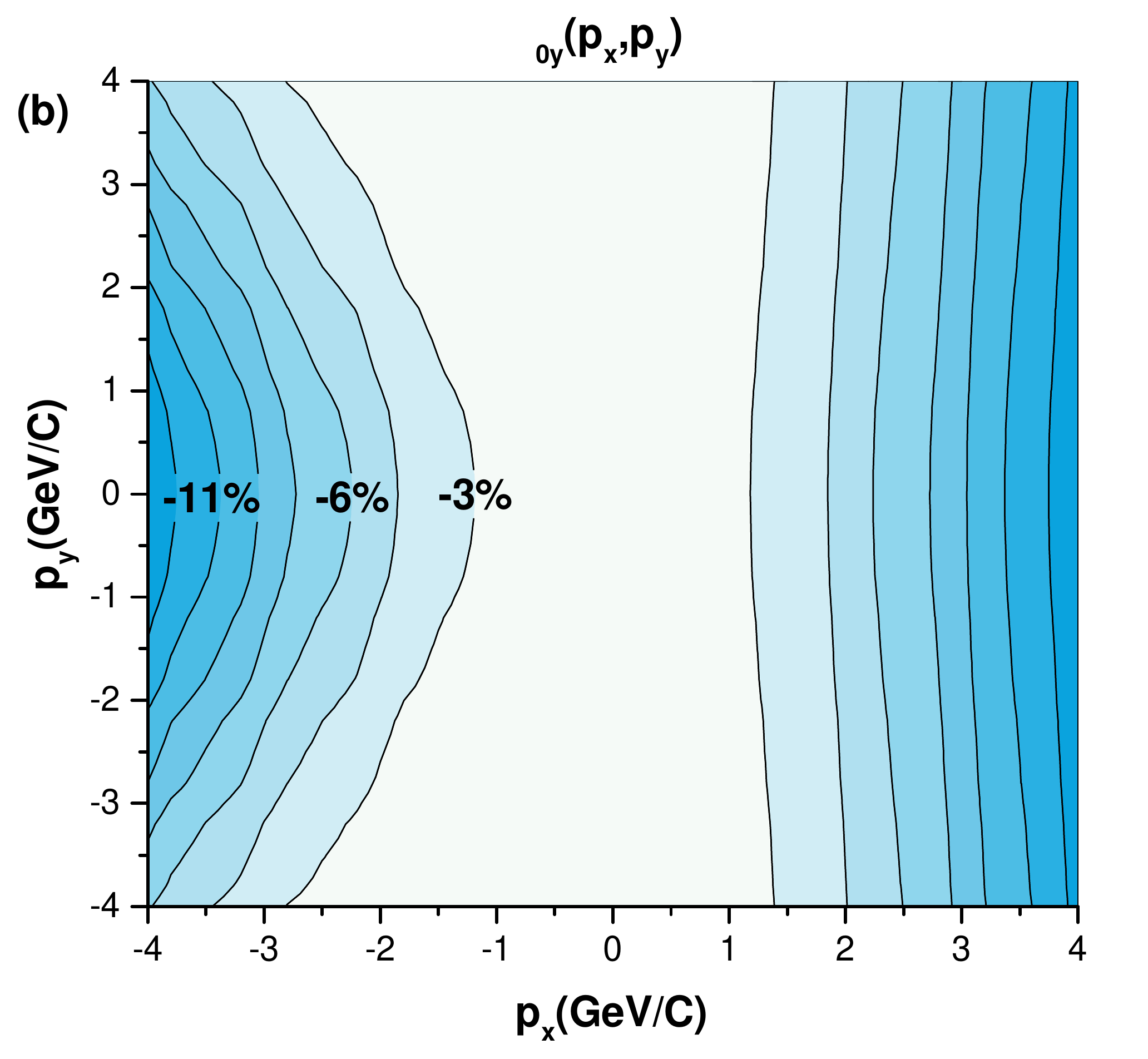}}
\caption{
(color online)
The (a) radial, $x$, and (b) axial, $y$, components of 
$\Lambda$-polarization, $\vec \Pi_0(\vec p)$, 
in the $\Lambda$'s rest frame. For $\Pi_{0x}(\vec p)$ the
contours represent changes of $1$\% from $-9.5$\% in the upper
left-hand corner to $9.5$\% in the upper right-hand corner, 
whereas
the contours of $\Pi_{0y}(\vec p)$ change in steps of $2$\% ranging
from $\Pi_{0y}=0$ (!) at the CM momentum ($p_x=p_y=0$) 
to $-12$\% for $p_x=\pm 4$GeV/c at the edges. 
Both plots are asymmetric due to the Lorentz boost to the
$\Lambda$ rest frame.
}
\label{F-P0xy}
\end{center}
\end{figure}
%%%%%%%%%%%%%%%%%%%%%%%%%%%%%%%%%%%

\end{widetext}

Now if we study the axis directed components, this is given by
$ \Pi_y =  \Pi_{1y} + \Pi_{2y}$. Both these terms have a 
negative 
maxima of the same magnitude, $-8$\%,  at the corners, $p_x,p_y=\pm 4$GeV/c,
thus these terms add up constructively and result in $\Lambda$-particle 
polarizations reaching -16\% at high momenta. At small momenta the polarization
is still the same sign but has a reduced value of the order of 1.5\% arising
from the vorticity (Fig. \ref{F-Py_xy}).

In this Exact model
the $x$ and $z$ components of the polarization arise only from the 
second term, $\vec \Pi_2(\vec p)$. The $x$ component
is reaching $\pm 8$\%, while the $z$ component is smaller, it reaches about
$\pm 3$\%. These both are asymmetric for $\pm p_y$ change, and show an 
opposite symmetry. 
The $x$-component is proportional to $p_y$ and the
dynamics of radial expansion. Thus it follows the signature of $p_y$,
Fig. \ref{F-P2x_xy}.
The $z$-component is proportional to $p_y$ and the
dynamics of radial expansion, thus it follows the signature of $p_y$,
Fig. \ref{F-P2z_xy}. The $z$-component is proportional to
$p_x \dot{\beta}_y$ and inversely proportional to
$p_y \dot{\beta}_x$, Fig. \ref{F-P2z_xy}. 
These two effects compensate each other so the 
maxima of the polarization are smaller and the symmetry is opposite 
to that of the $x$-component. This term is sensitive to the
balance between the axial expansion and the radial expansion in the model.

The $\Lambda$ polarization is measured via the 
angular distribution of the decay protons in the $\Lambda$'s rest 
frame, as shown in Eq. \ref{Pi0}. The resulting distribution is shown in
Fig. \ref{F-P0xy}. This new study indicates that the dynamics of the 
expansion may lead to non-negligible contribution to the observable 
polarization.
The structure of $\Pi_{0y}(\vec p)$ is similar to the one obtained in
Ref. \cite{BCsW13}, but here the contribution of the "second", 
$\partial_t \vec \beta$ term is also included, which makes the $y$-directed
polarization stronger at high $p_x$ values, 12\%,  while it was 9\% in 
Ref. \cite{BCsW13}, both in the negative  $y$-direction.  Furthermore,
the second term changes the structure, of the momentum dependence of
$\Pi_{0y}(\vec p)$, and it becomes $\pm p_x$ asymmetric. 

Recently the vorticity and polarization were also studied in two fluid 
dynamical models \cite{Bea15}. The initial states that were used 
from Bozek and Gubser  neglected fully the initial shear flow in the
central domain of the reaction, in contrast to other models where this 
is present \cite{M1,M2,CF13,GBL14,KFGY15}. 
This results in negligible thermal vorticity 
in the central domain of the collision (Figs. 3, 13 of Ref. \cite{Bea15}),
and consequently a negligible polarization from the vorticity from the 
"first term" discussed here. Thus, the observed vorticity arises from 
the "second term".  

On the other hand there is qualitative agreement between 
Figs. 12 of Ref. \cite{Bea15} and this work in the sense that only the
$y$-directed (i.e. [x,z] or [x,$\eta$]) component of the vorticity
leads to an overall average net polarization. This arises in both models
from the initial angular momentum and points into the $-y$-direction.
In Ref. \cite{Bea15} this arises as a consequence of viscous evolution
of the initial, vorticity-less flow, while in our Exact model it is
present in the initial state. 

Recent preliminary experimental results reported for the first time
\cite{Lisa},
significant $\Lambda$ and $\bar{\Lambda}$ polarization for
peripheral collisions at RHIC
for beam energies $\sqrt{s_{NN} }= 7.7 - 39$ GeV aligned with the
axis direction of the angular momentum of the participant system. 
Furthermore the $\Lambda$ and $\bar{\Lambda}$ polarizations were 
pointing in the same direction confirming our approach. 

In this work we analyzed and compared the two terms of polarization,
in the Exact model. Including both rotation and expansion, and vorticity
arising from both of these effects enables us to study the consequences
of the two terms separately.  This study indicates that the assumptions 
regarding the initial state are influencing the predictions on the 
observed vorticity, while in all cases observable polarization is predicted.

%%%%%%%%%%%%%%%%%%%%%%%%%%%%%%%%%%%
\section*{Acknowledgements}
%%%%%%%%%%%%%%%%%%%%%%%%%%%%%%%%%%

Enlightening discussions with Sharareh Mehrabi Pari, Francesco Becattini, 
Eirik Hatlen, Istvan Papp, Stuart Holland and Sindre Velle are gratefully 
acknowledged. One of the authors, Y.L. Xie, is supported by the China 
Scholarship Council.

%%%%%%%%%%%%%%%%%%%%%%%%%%%%%%%%%%%%%%%


\begin{thebibliography}{99}
%%%%%%%%%%%%%%%%%%%%%%%%%%%%%%%%%%%%%%%

\bibitem{M1}
%Initial state of ultrare-lativistic heavy ion collisions
V.K. Magas, L.P. Csernai, and D.D. Strottman,
 Phys. Rev. C {\bf 64} 014901 (2001).

\bibitem{M2}
%Effective string rope model for the initial stages of ultra-relativistic heavy ion collisions
V.K. Magas, L.P. Csernai, and D.D. Strottman,
 Nucl. Phys. A {\bf 712} 167 (2002).


\bibitem{McI-Teo}
 %Generalized planar black holes and the holography of hydrodynamic shear
 B. McInnes, and E. Teo,
 Nucl. Phys. B {\bf 878}, 186 (2014).

\bibitem{McI2014}
%The Limits of Gauge-Gravity Duality
B. McInnes,
Nucl. Phys. B {\bf 887}, 246 (2014).
% arXiv: 1403.3258v1 [hep-th].


\bibitem{Nona15}
K. Okamoto, C. Nonaka, and Y. Akamatsu,
% Kelvin-Helmholtz instability in relativistic heavy ion collisions,
Poster, presented at: The XXV International Conference on 
Ultrarelativistic Nucleus-Nucleus Collisions,
Sept. 27 - Oct. 3, 2015, Kobe, Japan; (published electronically).


\bibitem{Pang15}
L.G. Pang, G.-Y. Qin, V. Roy, X.-N. Wang, and G.-L. Ma,
% De-correlation of anisotropic flowalong the longitudinal direction,
Invited talk presented at The XXV International Conference 
on Ultrarelativistic Nucleus-Nucleus Collisions,
Sept. 27 - Oct. 3, 2015, Kobe, Japan; (published electronically).


\bibitem{CGSTNK82}
L.P. Csernai, W. Greiner, H. St{\"o}cker, I. Tanihata, 
S. Nagamiya, and J. Knoll, 
%Macroscopic nucleon-nucleon correlations caused by the bounce - off
%process in energetic collisions of heavy nuclei;
Phys. Rev. C {\bf 25}, 2482  (1982).

\bibitem{CFR84}
L.P. Csernai, G. Fai, and J. Randrup,
%Effect of fluctuations on global analysis of fluid-dynamical
%calculation for high energy nuclear collisions;
Phys. Lett. B {\bf 140}, 149  (1984).

\bibitem{hydro1}
L.P. Csernai, V.K. Magas, H. St\"ocker, and D.D. Strottman,
Phys. Rev. C {\bf 84},  024914 (2011).

\bibitem{hydro2}
L.P. Csernai, D.D. Strottman, and Cs. Anderlik,
Phys. Rev. C {\bf 85}, 054901 (2012).

\bibitem{CsN14}
T. Cs\"org{\H o}, and M.I. Nagy,
Phys. Rev. C {\bf 89}, 044901 (2014).
	
\bibitem{CsWCs14-2}
L.P. Csernai, D.J. Wang, and T. Cs\"org{\H o},
% A new family of exact and rotating solutions of fireball hydrodynamics
Phys. Rev. C {\bf 90}, 024901 (2014).

\bibitem{CsV13}
Csernai, L.P., and S. Velle,
%"Differential HBT Method to Analyze Rotation." 
Int. J. Mod. Phys. E {\bf 23}, 1450043 (2014).

\bibitem{CKM}
L.P. Csernai, J.I. Kapusta, and L.D. McLerran,
Phys. Rev. Lett. {\bf 97},  152303 (2006).

\bibitem{CMW13}
L.P. Csernai, V.K. Magas, and D.J. Wang,  
Phys. Rev. C {\bf 87}, 034906 (2013).

\bibitem{WNC13}
D.J. Wang, Z. N\'eda, and L.P. Csernai,
Phys. Rev. C {\bf 87}, 024908 (2013).

\bibitem{GBL14}
G. Graef, M. Bleicher, and M. Lisa,
Phys. Rev. C {\bf 89}, 014903 (2014).

\bibitem{BCsW13}
F. Becattini, L.P. Csernai, and D.J. Wang,
Phys. Rev. C {\bf 88}, 034905 (2013).

\bibitem{CsI15}
L.P. Csernai, and J.H. Inderhaug,
Int. J. Mod. Phys. E {\bf 24}, 1550013 (2015).
%arXiv:1503.03247v1 [nucl-th]
% DOI: 10.1142/S021830131550035

\bibitem{DM81}
T.A. DeGrand, and H.I. Miettinen, Phys. Rev. D {\bf 24}, 2419 (1981).
% forward production in small-transverse-momentum fragmentation
% processes  up to 30% !!!

\bibitem{LW05}
Z.-T. Liang, and X.-N. Wang, Phys. Lett. B {\bf 629}, 20 (2005);
Z.-T. Liang, and X.-N. Wang, Phys. Rev. Lett. {\bf 94}, 102301 (2005).
% spin-orbit interaction,   hadronization scenarios,
% showed that the results are depend on the hadronization
% mechanism,  investigated the effect of the decay products on the 
% v_2 flow harmonics.

\bibitem{ACHM02}
A. Ayala, E. Cuautle, G. Herrera, and L.M. Montano, 
Phys. Rev. C {\bf 65}, 024902 (2002).
% L production depende on QGP formation, If QGP then coalescence of 
% free valence quarks

\bibitem{BeChZaGr13}
F. Becatinni, V. Chandra, L. Del Zanna, and E. Grossi,
% Relativistic distribution function for particles with
% spin at local thermodynamical equilibrium
Annals of Physics {\bf 338}, 32 (2013).

\bibitem{CsK92}
% Nucleation in relativistic first-order phase transitions,
    L.P. Csernai, and J.I. Kapusta,
    Phys. Rev. {\bf D46} (1992) 1379-1390;
%    Dynamics of the QCD phase transition,
    L.P. Csernai, and J.I. Kapusta,
    Phys. Rev. Lett. {\bf 69}, 737 (1992).

\bibitem{CC94}
% Quark-gluon plasma freeze-out from a supercooled state?
    T. Cs\"org\H o, and L.P. Csernai,
    Phys. Lett. B {\bf 333}, 494 (1994). 

\bibitem{CM95}
%      Fast Hadronization of Supercooled Quark-gluon Plasma,
      L.P. Csernai, and I.N. Mishustin,
      Phys. Rev. Lett. {\bf 74}, 5005 (1995).

\bibitem{AlZ07}
I.S. Gradstein, and I.M. Ryzhik,
{\it Table of integrals, series, and products} (Academic Press, 2007).
%I.S. Gradshteyn and I.M. Ryzhik,
%Table of Integrals ,Series, and Products, Seven Edition
%\bibitem{IGF}
%  M. Abramowitz, and I.A. Stegun: {\it Handbook of mathematical functions}
%(Dover, New York, 1965) 6.5.2;
%I.S. Gradstein, and I.M. Ryzhik: {\it Table of Integrals ...},
%(Academic Press, 1994)

\bibitem{CsVW14}
L.P. Csernai, S. Velle, and  D.J. Wang,
% New method to detect rotation in high-energy heavy-ion collisions
Phys. Rev. C {\bf 89}, 034916 (2014).

\bibitem{Stoecker_handbook}
Horst St\"ocker: {\it Taschenbuch Der Physik},
(Harri Deutsch, 2000), 1.3.2/6d.

\bibitem{ChengEtAl10}
%   Matching stages of heavy ion collision models,
    Yun Cheng, L.P. Csernai, V.K. Magas, B.R. Schlei, and D. Strottman,
    Phys. Rev. C {\bf 81},  064910 (2010).
%    (arXiv: 1006.5820 v1 [nucl-th])

\bibitem{Lisa}
M.A. Lisa, invited talk at the XI Workshop on Particle Correlations 
and Femtoscopy (WPCF2015), 3-7 November 2015, Warsaw, Poland.

\bibitem{Bea15}
F. Becattini, G. Inghirami, V. Rolando, A. Beraudo,
L. Del Zanna, A. De Pace, M. Nardi, G. Pagliara, and V. Chandra,
%A study of vorticity formation in high energy nuclear collisions
Eur. Phys. J. C {\bf 75}, 406 (2015).
%arXiv:1501.04468v2 [nucl-th] (2015).

\bibitem{CF13}
G.-Y. Chen, and R.J. Fries,
%Global flow of glasma in high energy nuclear collisions
Phys. Lett. B {\bf 723}, 417 (2013).

\bibitem{KFGY15}
%Early Time Dynamics of Gluon Fields in High Energy
%Nuclear Collisions,
J. Kapusta, R.J. Fries, G.-Y. Chen, and L. Yang,
Invited talk presented at The XXV International Conference 
on Ultrarelativistic Nucleus-Nucleus Collisions,
Sept. 27 - Oct. 3, 2015, Kobe, Japan; (published electronically).


%\bibitem{ACLS01}
%Simple solutions of fireball hydrodynamics for self-similar elliptic flows
%S. V. Akkelin, T. Csorgo, B. Lukacs, Yu. M. Sinyukov, M. Weiner,
%Phys. Lett. B {\bf 505}, 64 (2001), arXiv: hep-ph/0012127;
%
%Simple solutions of fireball hydrodynamics for self-similar, ellipsoidal flows
% T. Cs\"orgo,  Acta Phys. Polon. B 37, 483 (2006).
%arXiv: hep-ph/0111139

%\bibitem{RungeKutta} W.E. Boyce and R.C. DiParma: {\it
%Elementary Differential Equations and Boundary Value Problems},
%(Wiley, 1997).

%\bibitem{CsCsHK03}
%T. Cs\"org{\H o}, L.P. Csernai, Y. Hama, T. Kodama,
%% Simple solutions of relativistic hydrodynamics
%% for systems with ellipsoidal symmetry, 
%Heavy Ion Phys. {\bf A21} 73-84, (2004). arXiv: 0306.004v1 [nucl-th]

%\bibitem{BCKL85}
%H.W. Barz, L.P. Csernai, B. K\"ampfer, B. Luk\'acs
%Stability of detonation fronts leading to quark-gluon plasma
%Phys. Rev. D {\bf 32}, 115  (1985).




\end{thebibliography}
\end{document}